\documentclass[acmsmall]{acmart}

\usepackage{graphicx}
\usepackage{balance}  
\usepackage{subfig}
\usepackage{float}
\usepackage{bbding}
\usepackage{url}
\usepackage{bm}

\usepackage{color,xcolor}
\usepackage{algorithm}
\usepackage[noend]{algpseudocode}

\newcommand{\revision}[1]{\textcolor{black}{#1}}

\usepackage{xspace}
\newcommand{\kw}[1]{{\ensuremath {\mathsf{#1}}}\xspace}

\newcommand{\sysname}{\kw{HongTu\xspace}}
\newcommand{\sysnameb}{{\ensuremath {\mathsf{\bm{HongTu}}}}\xspace}
\usepackage{amsmath}
\usepackage{graphics}
\usepackage{epsfig}
\usepackage{enumitem}
\usepackage{libertine}
\usepackage{pifont}
\newcommand{\Paragraph} [1] {\smallskip\noindent{\bf #1. }}
\usepackage{booktabs}
\usepackage{multirow}
\usepackage[misc]{ifsym} 
\usepackage[font=small,labelfont=bf]{caption}
\usepackage{hyperref}

\usepackage{tcolorbox}


\setcopyright{rightsretained}
\acmJournal{PACMMOD}
\acmYear{2023} \acmVolume{1} \acmNumber{4 (SIGMOD)} \acmArticle{246} \acmMonth{12} \acmPrice{}\acmDOI{10.1145/3626733}

\makeatletter
\gdef\@copyrightpermission{
  \begin{minipage}{0.2\columnwidth}
   \href{https://creativecommons.org/licenses/by/4.0/}{\includegraphics[width=0.90\textwidth]{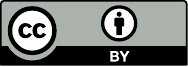}}
  \end{minipage}\hfill
  \begin{minipage}{0.8\columnwidth}
   \href{https://creativecommons.org/licenses/by/4.0/}{This work is licensed under a Creative Commons Attribution International 4.0 License.}
  \end{minipage}
  \vspace{5pt}
}
\makeatother

\received{April 2023}
\received[revised]{July 2023}
\received[accepted]{August 2023}

\begin{document}



\title{HongTu: Scalable Full-Graph GNN Training on Multiple GPUs (via communication-optimized CPU data offloading)}

\author{Qiange Wang}
\affiliation{%
  \institution{National University of Singapore\country{Singapore}}
  }
\email{wangqg@comp.nus.edu.sg}
\author{Yao Chen}
\affiliation{%
  \institution{National University of Singapore\country{Singapore}}
  }
\email{yaochen@comp.nus.edu.sg}

\author{Weng-Fai Wong}
\affiliation{%
  \institution{National University of Singapore\country{Singapore}}
  }
\email{wongwf@comp.nus.edu.sg}

\author{Bingsheng He}
\affiliation{%
  \institution{National University of Singapore\country{Singapore}}
  }
\email{hebs@comp.nus.edu.sg}


\begin{abstract}
Full-graph training on graph neural networks (GNN) has emerged as a promising training method for its effectiveness. Full-graph training requires extensive memory and computation resources. To accelerate this training process, researchers have proposed employing multi-GPU processing. However the scalability of existing frameworks is limited as they necessitate maintaining the training data for every layer in GPU memory. To efficiently train on large graphs, we present \sysname, a scalable full-graph GNN training system running on GPU-accelerated platforms. \sysname stores vertex data in CPU memory and offloads training to GPUs. \sysname employs a memory-efficient full-graph training framework that reduces runtime memory consumption by using partition-based training and recomputation-caching-hybrid intermediate data management. To address the issue of increased host-GPU communication caused by duplicated neighbor access among partitions, \sysname employs a deduplicated communication framework that converts the redundant host-GPU communication to efficient inter/intra-GPU data access. Further, \sysname uses a cost model-guided graph reorganization method to minimize communication overhead.
Experimental results on a 4$\times$A100 GPU server show that \sysname effectively supports billion-scale full-graph GNN training while reducing host-GPU data communication by 25\%-71\%. Compared to the full-graph GNN system DistGNN running on 16 CPU nodes, \sysname achieves speedups ranging from 7.8$\times$ to 20.2$\times$. For small graphs where the training data fits into the GPUs, \sysname achieves performance comparable to existing GPU-based GNN systems.





\end{abstract}
\maketitle



\section{Introduction}
\label{sec:intro}

Graph neural networks (GNNs) have gained increasing popularity for their effectiveness in modeling graph data \cite{GGSNN_ICLR_2016, HGNN_CoRR_2019, CNNonGraph_NIPS_2016, ESGCN_EMNLP_2017, GCN_ICLR_2017, GIN_ICLR_2019, gnn_survey_arxiv2019, GAT_ICLR_2018, GNNRS_CoRR_2020,LargeEA22,DualMatch}. By iteratively aggregating parameterized neighbor representations through graph propagation, GNNs can capture the topology and feature information at the same time and generate more informative representations for the downstream tasks.

Recently, full-graph GNN training that trains on the entire graph, has emerged as a promising GNN training method for its effectiveness brought by full-neighbor aggregation semantic and full-batch gradient descent \cite{ROC,DISTGNN_ARXIV_2021,NEUTRONSTAR_SIGMOD_2022}. However, full-graph training requires high computing power. The training involves random vertex data access and neural network computation, requiring high memory bandwidth and massive parallel computation. Considering the increasing sizes of real-world graphs, research efforts have been made toward extending GNN training to GPU platforms \cite{ROC,SC20,NEUGRAPH_ATC_2019,NEUTRONSTAR_SIGMOD_2022,PIPEGCN_ICLR_2022,DGCL_EUROSYS_2021,SANCUS_VLDB_2022}. These frameworks partition the input graphs, parallelize the computation across multiple GPUs, and handle remote neighbor aggregations through inter-GPU communication.

\begin{table}
	\vspace{-0.05in}
	\caption{Memory consumption of graph topology, vertex (\textbf{Vtx}) data, and intermediate (\textbf{Intr}) data for 3-layer full-graph GCN training on three billion-scale graphs.}
	\vspace{-0.1in}
	\label{tab:sec2:mem}
	\centering
	\small
	{\renewcommand{\arraystretch}{1.2}
	\begin{tabular}{l c c c c c c c}
		\hline
		
		\hline
		{\textbf{Dataset}} &
		{\textbf{Model Config}}  &{\textbf{Topology}}&
		{\textbf{Vtx Data}}&\textbf{Intr Data}\\
		\hline
        {it-2004}  &256-128-128-64& 12.8GB&177.2GB&108.3GB\\
		{ogbn-paper}  &200-128-128-172& 18.0GB&519.4GB&425.3GB\\

        {friendster} &256-128-128-64& 28.9GB&293.3GB&179.3GB\\

		\hline

	\end{tabular}
	}
	\vspace{-0.15in}
\end{table}

Despite the significant performance improvement achieved through massive parallel processing, scaling GPU-based GNN training frameworks to large graphs remains challenging due to the limited capacity of GPU memory. In GNN training, the core data that need to be maintained in GPU memory includes vertex data and intermediate data. Vertex data consist of the vertex representations (feature) and vertex gradients of every layer, while intermediate data are the intermediate results of neural network models, which are generated in the forward computation and consumed in the gradient computation in the backward pass. For real-world graphs, the data often exceed the device memory capacity even with multiple GPUs. As illustrated in Table \ref{tab:sec2:mem}, both the vertex data and intermediate data can occupy hundreds of gigabytes of GPU memory\footnote{The memory consumption of intermediate data varies across GNN models and can be much larger in GNNs involving complex edge computation \cite{GGCN_ICLR_2016,GAT_ICLR_2018}.}. Moreover, in practice, vertex data and intermediate data are only the basic data for model training, and additional memory must be reserved for auxiliary data such as graph topology, communication buffers, and neighbor replicas. The experiments in DistGNN \cite{DISTGNN_ARXIV_2021} demonstrate that running a 3-layer GraphSAGE model on the ogbn-paper graph \cite{OGB_DATASET}, with 111 million vertices and 1.6 billion edges, requires 6 terabytes of memory in a cluster with 16 shared-nothing CPU nodes \cite{DISTGNN_ARXIV_2021}. Building a GPU memory pool of comparable size requires significant monetary cost and engineering effort, which limits the graph scale that existing full-graph GNN systems can handle.

Recently, mini-batch GNN training has been proposed as an alternative approach to training large graphs on memory-constrained platforms. This method enables users to train on batched and sampled subgraphs, thus reducing the memory requirements \cite{aligraph_vldb2019,DORYLUS_OSDI_2021,GNNLAB_EUROSYS_2022,PaGraph_SoCC_2020,DISTDGL_ARXIV_20,Euler}. However, mini-batch training may suffer from low accuracy caused by information loss \cite{DORYLUS_OSDI_2021,ROC,DISTGNN_ARXIV_2021,NEUTRONSTAR_SIGMOD_2022}.

In this work, we aim to accelerate large-scale full-graph GNN training beyond GPU memory capacity by leveraging partition-based CPU-GPU heterogeneous processing, which partitions and stores data in CPU memory and offloads computation to GPUs.
This approach has been widely adopted in GPU applications that manage data beyond GPU memory capacity, including database analytical processing  \cite{GRDBMS_SIGMOD_2015,GRDBMS_VLDB_2016} and graph data analytics \cite{EMOGI_VLDB_20,graphreduce_sc_2015}. However, traditional partition-based processing approaches face two significant challenges when handling neural network computation and high-dimensional graph propagation in full-graph GNN training.

\Paragraph{Firstly, partition-based processing cannot effectively reduce memory consumption of intermediate data} 
In DNN training, data samples can be randomly partitioned into disjoint subsets for parallel training, which reduces the memory consumption of both vertex and intermediate data since there are no dependencies among the data samples. However, GNN training is distinct from it because the graph propagation computation creates cross-partition data dependencies. To compute the gradients of a partitioned subgraph, gradients of all its dependent partitions from downstream layers must also be computed \cite{NEUTRONSTAR_SIGMOD_2022}, which in turn requires storing intermediate data for these partitions. As a result, although partition-based processing enables loading the vertex data of a small partitioned slice at a time, a significant amount of GPU memory still needs to be reserved for maintaining the intermediate data. 

\vspace{-0.03in}
\Paragraph{Secondly, partition-based processing leads to increased host-GPU communication} In GNN training, graph propagation involves aggregating features from neighboring vertices. This requires loading the data of the entire neighbor set onto GPUs when processing each partitioned subgraph.
However, when a large graph is split into multiple partitions, vertices with multiple outgoing edges may be replicated to multiple partitions as neighbors. As a result, it is necessary to transfer them multiple times during training, which increases host-GPU communication.
Moreover, since high-dimensional vertex attributes can consume a substantial amount of memory (as shown in Figure \ref{tab:sec2:mem}), it is not feasible to store the frequently accessed vertex data entirely in GPU, as is done in GPU-accelerated graph analytical frameworks \cite{EMOGI_VLDB_20,scaph_atc_2020,subway_eurosys_2020}.


We present \sysname, a GPU-accelerated full-graph GNN training system that addresses the challenges of traditional partition-based processing through two critical functions. Firstly, \sysname employs a \textbf{memory-efficient GNN training framework} that reduces runtime memory consumption of both vertex and intermediate data. This framework integrates a GNN-friendly partition method and a cost-effective recomputation-caching-hybrid intermediate data management method. Inspired by the recomputation-based DNN training method that avoids storing intermediate data by releasing intermediate data in the forward pass and recomputing it in the backward pass\cite{SUBMEM_ARXIV_2016}. Based on the original method, our recomputation-caching-hybrid method further combines GPU-based recomputation and CPU-based data caching to reduce the recomputation overhead in GNNs. Secondly, \sysname employs a \textbf{deduplicated communication framework} that reduces host-GPU communication for duplicated neighbor access among partitions. We observe that duplicated neighbors access between sequentially and concurrently scheduled subgraphs can be efficiently handled through a single host-GPU communication and multiple inter/intra-GPU data accesses, rather than communicating them individually between CPU and GPUs. We leverage this observation to develop a communication deduplication method, and we also propose a subgraph reorganization method that enhances the effect of communication deduplication to improve performance.

In summary, we make the following contributions. 

\vspace{-0.03in}
\begin{itemize}[leftmargin=*]
   \item  We propose a memory-efficient GNN training framework that reduces runtime memory consumption by integrating a partition-based GNN training method and a recomputation-caching-hybrid intermediate data management method.
   
   \item We propose a deduplicated communication framework that reduces host-GPU data communication by optimizing the duplicated neighbor accesses between sequentially and concurrently scheduled subgraphs.

   \item We develop \sysname, a GPU-accelerated system for full-graph GNN training that overcomes the memory limitation of GPUs and integrates an efficient communication implementation to achieve high performance.

\end{itemize}


Experimental results on four NVIDIA A100 GPUs show that \sysname reduces host-GPU communication by 38\%-78\% and achieves 1.3$\times$-3.4$\times$ performance improvement over the vanilla approach that transfers the entire neighbor set for each partition. When compared to DistGNN \cite{DISTGNN_ARXIV_2021} running on 16 CPU nodes, \sysname achieves speedups ranging from 7.8$\times$ to 20.2$\times$. \revision{Furthermore, for small graphs that can fit into GPUs, \sysname achieves performance comparable to existing multi-GPU systems.}

The rest of the work is organized as follows, §2 describes the background and motivations. §3 gives an overview of \sysname. §4 describes the memory-efficient GNN training framework. §5 describes the deduplicated communication framework. §6 describes system implementation. §7 presents results. §8 concludes.

\section{Background and Motivations}
\label{sec:2}

\subsection{Multi-GPU Architecture}
\begin{figure}
	\centering
 \vspace{-0.1in} 
	\includegraphics[width=2.8in]{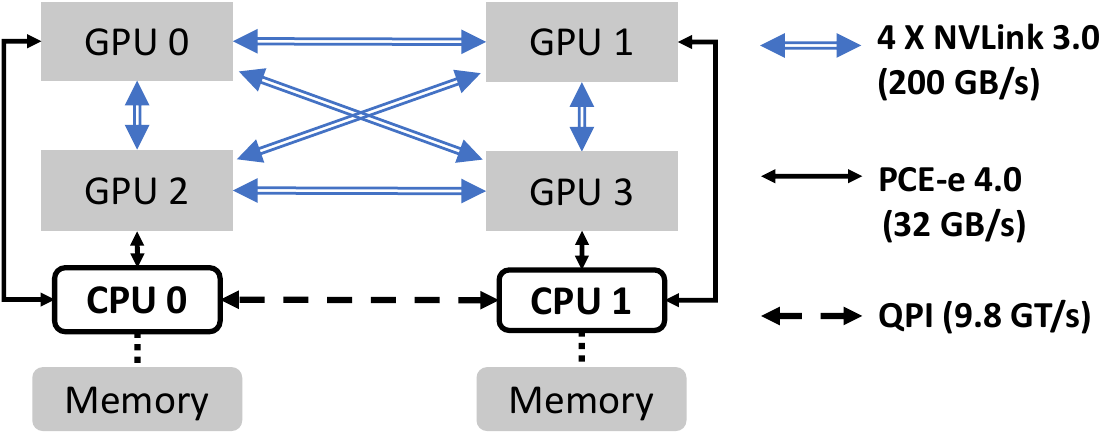}
	\caption{An example of multi-GPU architecture.}
    \vspace{-0.05in} 
	\label{fig:sec2:gpu_arch}
 \vspace{-0.06in}
\end{figure}

Modern GPUs are equipped with high bandwidth memory and massive streaming multiprocessors (SMs), making them suitable for memory- and computing-intensive applications. However, the limited device memory capacity, typically ranging from several to tens of gigabytes \cite{A100}, poses a constraint on the size of data that can be efficiently processed. To address this issue, hardware researchers have developed fast interconnects to connect multiple GPUs, such as AMD Infinity Fabric \cite{AMD_FAB} and NVIDIA NVLink \cite{NVLINK}. Figure \ref{fig:sec2:gpu_arch} provides an example of a 4$\times$A100 GPU server, where the four GPUs are interconnected through 4$\times$NVLink-3.0 with 200GB/s inter-GPU communication bandwidth, enabling low latency and high throughput inter-GPU data access. Every two GPUs are connected to a single CPU via PCIe 4.0 interconnect. Although CPUs are generally equipped with hundreds to thousands of gigabytes of host memory, the slow CPU-GPU communication bandwidth (up to 32GB/s in PCIe 4.0) often creates a performance bottleneck for GPU access to CPU memory. Moreover, the two CPUs are linked through a QPI bus, forming a two-socket Non-Uniform Memory Access (NUMA) architecture, where GPUs accessing remote CPU memory via QPI experience slower speeds than those accessing local CPU memory. Therefore, building high-performance multi-GPU applications requires careful optimization of heterogeneous communication, especially in reducing CPU-GPU data transfer. 




\vspace{-0.04in}
\subsection{GNN Basis}
\label{sec:basis}
A GNN takes a graph and the vertex-associated property (feature) of all vertices as input and learns a representation vector for each vertex by stacking multiple GNN layers. In each layer, GNN models generally follow an aggregate-update computation pattern.

\vspace{-0.1in}
\begin{align}\small
\label{eq:aggr}
\textbf{h}_v^l=\textbf{UPDATE}\Big(\textbf{AGGREGATE}(\{\textbf{h}^{l-1}_u| u\in N(v)\}), \textbf{h}^{l-1}_v\Big),
\end{align}
$\textbf{h}_{v}^l$ is the representation of $v$ in the $l$-th layer and $\textbf{h}_{v}^0$ is the input vertex feature. The \textbf{AGGREGATE}
function collects the $l$$-$$1$-th layer representations of $v$'s in neighbors, i.e., $\{\textbf{h}^{l-1}_u)| u\in N(v)\}$, to compute the neighbor representation of $v$. The \textbf{UPDATE} function utilizes the neighbor aggregation result and $v$’s representation in the $l$$-$$1$-th layer to calculate $v$’s vertex representation in the $l$-th layer. Both the aggregate and update functions can be neural networks, which are updated during training. To illustrate, we present two examples: the graph convolutional network (GCN) \cite{GCN_ICLR_2017} and the graph attention network (GAT) \cite{GAT_ICLR_2018}.

\noindent\textbf{GCN} is a simple yet effective model that has neural network computation on vertices.
\vspace{-0.05in}
\begin{align}\small
\label{eq:gcn}
\textbf{h}_v^l=\sigma(W^l\otimes(\sum_{_{u\in N(v)}} d_{uv} \textbf{h}^{l-1}_u))
\end{align}
The aggregate function is a simple weighted neighbor convolution, where $d_{uv}$ is the normalized edge weight of edge <$u,v$>. The {update} function involves a {linear} transformation and a non-linear activation function (e.g., $\textbf{ReLU}$).

\noindent\textbf{GAT} introduces a self-attention mechanism, which assigns different attention parameters on edges to distinguish which neighbors are more important.
\begin{align}\small
\label{eq:gat}
\textbf{h}_v^l=\sigma\Big(\sum_{_{u\in N(v)}}\frac{exp\big(\hat{\sigma}(a^l[W^l \textbf{h}^{l-1}_v ||W^l \textbf{h}^{l-1}_u])\big)}{\sum_{_{u\in N(v)}} exp\big(\hat{\sigma}(a^l[W^l \textbf{h}^{l-1}_v ||W^l \textbf{h}^{l-1}_u])\big)} \textbf{h}^{l-1}_u\Big)
\end{align}
 The aggregate function first concatenates the parameterized representations of source $u$ and destination $v$, and applies $a^l$ to compute the edge-wise attention coefficient, i.e., $a^l[W^l \textbf{h}^{l-1}_v ||W^l \textbf{h}^{l-1}_u]$. Then, it feeds the attention coefficients to a \textbf{LeakyReLU} activation (i.e., $\hat{\sigma}$) and uses a neighbor-oriented \textbf{softmax} function (i.e., $\frac{exp(\cdot)}{\sum_{_{u\in N(v)}} exp(\cdot)}$) to compute the edge weight for the neighbor aggregation. The {update} function is usually a simple non-linear activation (e.g., \textbf{ReLU}).

\begin{table}
\vspace{-0.1in}
	\caption{{Summary of existing full-graph GNN systems. The `VD' represents the vertex data and the `ID' represents the intermediate data.}}
	\vspace{-0.1in}
	\label{tab:system_overview}
	\centering
	\scriptsize
	{\renewcommand{\arraystretch}{1.2}
	\begin{tabular}{l | l |c |c |c |l}
		\hline
  
		\hline
		\multirow{2}*{\textbf{Hardware}}&\multirow{2}*{\textbf{Systems}} & \textbf{Full Nbr.}&
        {\textbf{Partially}} &{\textbf{Partially}}&
        \multirow{2}*{\textbf{Major contributions}}\\
        &&\textbf{Agg.}&{\textbf{ storing VD }}&{\textbf{storing ID}}&\\
		\hline

  \hline
  	\multirow{2}*{CPU}&{DistGNN}\cite{DISTGNN_ARXIV_2021}&\CheckmarkBold&-&- &CPU-aggregation optimization\& Staleness communication\\
   \cline{2-6}
		&{Graphite}\cite{GRAPHITE_ISCA_2022}&\CheckmarkBold& - &- &Hardware-assisted node property aggregation \\	\cline{2-4}	
		\hline

        \hline
	    \multirow{12}*{GPU}
     &{DGL}\cite{dgl_system}&\CheckmarkBold &\XSolidBrush&\XSolidBrush & single-GPU system \\
     \cline{2-6}
     
     &{PyG}\cite{PYG_ARXIV_2019}&\CheckmarkBold &\XSolidBrush&\XSolidBrush & single-GPU system \\
     \cline{2-6}
    &{CAGNET}\cite{SC20}&\CheckmarkBold &\XSolidBrush&\XSolidBrush & 1.5D/2D/3D Graph Partitioning \\
     \cline{2-6}

    &{DGCL}\cite{DGCL_EUROSYS_2021}& \CheckmarkBold&\XSolidBrush & \XSolidBrush& Cost-based communication routine \\
     \cline{2-6}
     &{PipeGCN}\cite{PIPEGCN_ICLR_2022}&\XSolidBrush& \XSolidBrush&\XSolidBrush & Staleness-communication\&Pipelining   \\
	\cline{2-6}
 &{Sancus}\cite{SANCUS_VLDB_2022}&\CheckmarkBold& \XSolidBrush & \XSolidBrush& Staleness-communication  \\
        \cline{2-6}

        \cline{2-6}
        &{NeuGraph}\cite{NEUGRAPH_ATC_2019}&\XSolidBrush&{\CheckmarkBold} & \XSolidBrush& SAGA abstraction \& Partition-based training \\
            \cline{2-6}&{NeutronStar}\cite{NEUGRAPH_ATC_2019}&\XSolidBrush& {\CheckmarkBold}& \XSolidBrush& Hybrid dependency management\&Partition-based training    \\
	\cline{2-6}
        
        &\multirow{2}*{{ROC}\cite{ROC} }&\multirow{2}*{\CheckmarkBold}&\multirow{2}*{\XSolidBrush}  &\multirow{2}*{\CheckmarkBold} &  Learned graph partitioning  \\
        &&&& &  cost-based intermediate data management   \\
        
        \cline{2-6}
        &\multirow{2}*{\sysnameb}&\multirow{2}*{\CheckmarkBold}&\multirow{2}*{\CheckmarkBold}  &\multirow{2}*{\CheckmarkBold} &  Recomputation-caching-hybrid intermediate data management  \\
        &&&& &  Deduplicated communication framework   \\
		\hline
  
		\hline
	\end{tabular}
	}
 \vspace{-0.05in}
\end{table}

\subsection{Full-Graph GNN Training}
Full-graph GNN training uses the full-neighbor aggregation semantic and global gradient descent algorithm. It runs epochs repeatedly on the entire graph until reaching the target accuracy or epoch. Each training epoch consists of a forward and a backward pass, followed by parameter update, which uses the gradients computed in the backward pass to update the trainable parameters in every layer. \revision{In the forward pass, vertex representations are computed layer-by-layer using the \textbf{AGGREGATE} and \textbf{UPDATE} operations presented in Section \ref{sec:basis}. At every layer, each vertex aggregates the representations of the incoming neighbors and calculates the vertex representation by applying the learnable model parameters. The final layer's vertex representations are then sent to the downstream task where the loss value is calculated based on the ground truth labels. In the backward pass, GNN's computation starts from the last layer and proceeds back to the first layer, calculating the gradient of loss with respect to the model across all layers. In each layer, the gradients of vertex representations are computed using the chain rule, facilitating both intra-layer model gradient calculation and cross-layer gradient transmission \cite{NEUTRONSTAR_SIGMOD_2022,DORYLUS_OSDI_2021}.} 

\revision{GNN models are distinct from traditional DNN models because the link relationship between vertices creates complex and non-uniform data dependencies. In a CNN, the convolution kernels are fixed and treat all pixels in the same way. However, in a GCN, the \textbf{AGGREGATE} and its backward operation handle data dependencies by gathering data along edges. This not only entails random data accesses but also introduces complexities in workload partition due to its irregular nature. Generally, achieving efficient vertex data access necessitates accommodating all vertex data within GPU memory.} In GNN training, intermediate data generated in the forward pass needs to be reserved for gradient computation in the backward pass. For example, the update function in GCN \cite{GCN_ICLR_2017} involves \texttt{linear}+\texttt{Relu} computations in the forward pass, i.e., $\textbf{h}=ReLU(\textbf{a}\times W)$. Its backward pass computes the gradients of parameter $W$ using the formula: $\nabla W$=$(\textbf{a})^t\times ReLU^{-1}(\textbf{a}\times W)*\nabla\textbf{h}$. Here, $ReLU^{-1}$ is the derivative function of $ReLU$, which returns 1 for positive inputs and 0 otherwise. $(\cdot)^t$ represents the transpose operation. $\textbf{a}\times W$ is the intermediate data that needs to be reserved for gradient computation. While the effectiveness and high accuracy of full-graph GNN training have been widely demonstrated by academic studies \cite{NEUGRAPH_ATC_2019,ROC,DORYLUS_OSDI_2021,GCN_ICLR_2017}, its practical application in industry is limited due to the significant memory requirements for maintaining large-scale vertex and intermediate data.


\vspace{-0.1in}
\subsection{Existing Systems and limitations} 
\label{sec2:frame}

Table \ref{tab:system_overview} summarizes existing full-graph GNN systems and their major contributions. 
\revision{Early systems, such as DGL \cite{DGL_ARXIV_2019} and PyG \cite{PYG_ARXIV_2019} use full-graph training on a single GPU, and thus their efficiency and scalability are constrained by the limited GPU resource.}
To meet the high computation and memory requirements of full-graph GNN training,  distributed- and multi-GPU-based systems have been proposed. CAGNET \cite{SC20}, DGCL \cite{DGCL_EUROSYS_2021}, PipeGCN\cite{PIPEGCN_ICLR_2022}, and Sancus \cite{SANCUS_VLDB_2022} are four multi-GPU GNN systems that maintain both vertex and intermediate data in GPU memory. In these systems, inter-GPU communication emerges as a performance bottleneck \cite{SANCUS_VLDB_2022,DGCL_EUROSYS_2021}.
CAGNET \cite{SC20} proposes 1.5D, 2D, and 3D graph partitioning to optimize the data distribution among GPUs. DGCL \cite{DGCL_EUROSYS_2021} analyses the speeds of heterogeneous communication among devices and proposes an automatic routine algorithm to improve communication efficiency. PipeGCN \cite{PIPEGCN_ICLR_2022} and Sancus \cite{SANCUS_VLDB_2022} investigate staleness-communication in GNN training, which reduces communication times while sustaining a reasonable level of accuracy. Despite the high performance of these frameworks, they can hardly scale to large input graphs. As illustrated in Table \ref{tab:sec2:mem}, accommodating the data for training ogbn-paper graph needs at least 77 NVIDIA A100 GPUs (80GB), which is expensive and requires sophisticated design for managing communication and fault-tolerance on a distributed GPU cluster. Moreover, the relatively slow inter-node communication can also become a critical performance bottleneck \cite{DGCL_EUROSYS_2021}.


 \begin{figure}
 \vspace{-0.05in}
	\centering
	\includegraphics[width=0.58\linewidth]{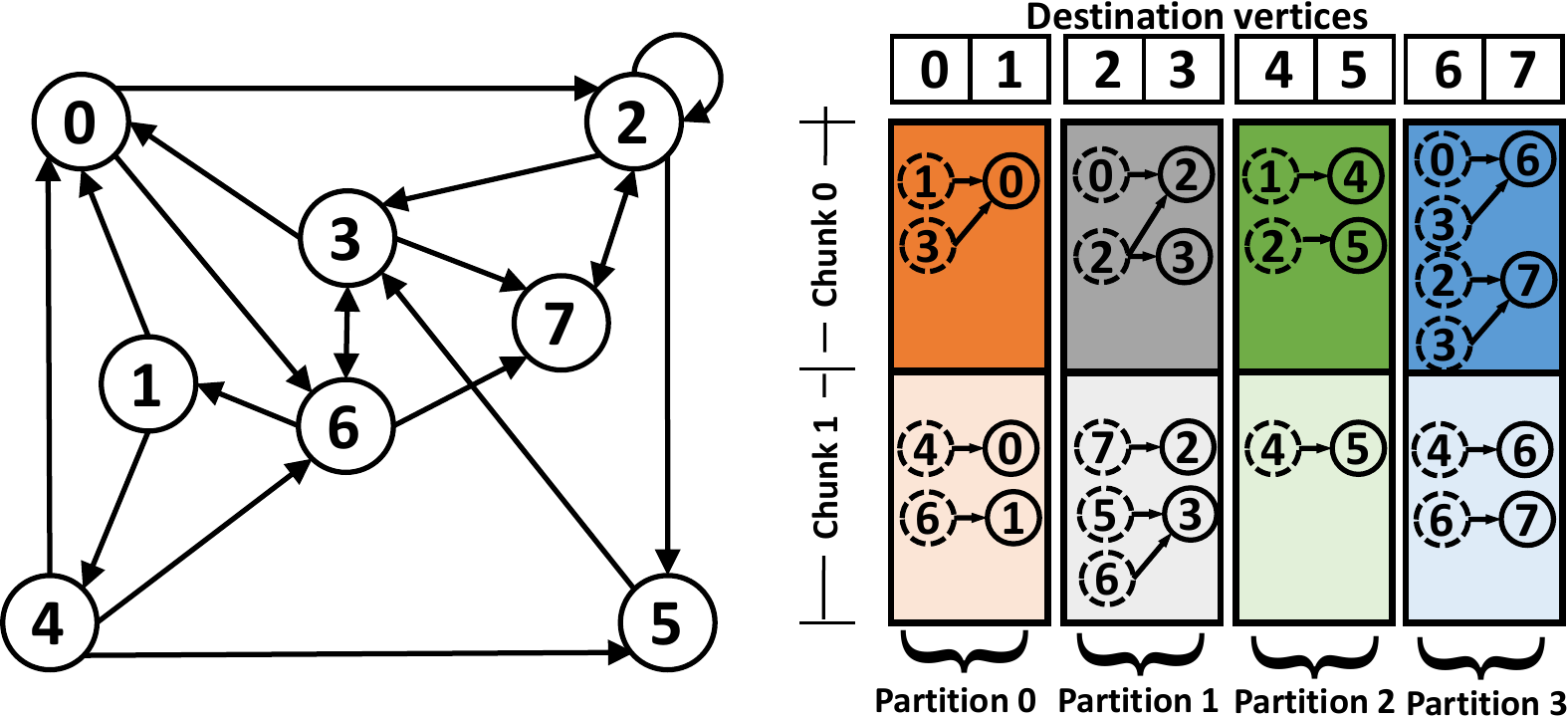}
 \vspace{-0.1in}
	\caption{2D graph partitioning on a 8 vertices toy graph. Each partitioned subgraph is represented by a colored box. Solid and dashed circles denote the master and mirror vertices, respectively.}
 	\label{fig:sec3:graph_partitioning}
\vspace{-0.1in}
\end{figure}

 Recently, some research work \cite{ROC,NEUGRAPH_ATC_2019,NEUTRONSTAR_SIGMOD_2022} try to relax the memory constraint of GPU by partially loading vertex or intermediate data during training. As illustrated in Table \ref{tab:system_overview}, NeuGraph \cite{NEUGRAPH_ATC_2019} and NeutronStar \cite{NEUTRONSTAR_SIGMOD_2022} employ 2-D graph partitioning to split a large graph into multiple chunks, where each chunk contains a specific range of destination and source vertices (as shown in Figure \ref{fig:sec3:graph_partitioning}). During training, the two frameworks store the vertex data in the CPU and sequentially load the vertex data of partitioned chunks to the GPU for training. On the other hand, ROC \cite{ROC} utilize CPU-memory to manage the intermediate data. ROC includes a cost model to represent the host-GPU data transfer overhead, and utilizes dynamic programming to find the optimal communication plan. By doing so, ROC allows the GPU to store only part of the intermediate data. 
 
 In addition to GPU-based systems, researchers have also explored distributed CPU-based systems \cite{DISTGNN_ARXIV_2021,GRAPHITE_ISCA_2022} to leverage the large memory capacity of CPU platforms. However, these systems generally exhibit inferior performance when compared to GPU-based solutions, and the monetary cost of using high-end CPU clusters is also high. Therefore, building a CPU-GPU heterogeneous system that fully utilizes the memory and computation resources of a single-node-multi-GPU architecture becomes a cost-effective option. 
 
 However, we observe that the existing out-of-GPU-memory processing systems still face two limitations that hinder their effectiveness and efficiency in handling large-scale GNN training.

\noindent{\textbf{{Limitation 1:}}} Existing systems still suffer from the high memory consumption of either vertex or intermediate data. While NeuGraph \cite{NEUGRAPH_ATC_2019} and NeutronStar \cite{NEUTRONSTAR_SIGMOD_2022} decrease the memory consumption of vertex data, they still require intermediate data to be stored entirely in the GPU. Conversely, ROC reduces the memory consumption of intermediate data, but still necessitates completed storage of vertex data in the GPU. More importantly, several critical limitations hinder the direct combination of these memory reduction methods. Firstly, the 2D partitioning in NeuGraph and NeutronStar separates a vertex's neighbors into multiple slices, making implementing full-neighbor aggregation challenging for complex GNNs like GAT \cite{GAT_ICLR_2018} model, which involves a {\texttt{softmax()}} computation on the entire neighbor set. In these workloads, loading all neighbor-containing partitions is still necessary. This renders existing systems ineffective on training large-scale GAT-like models. Secondly, ROC's caching-based method is inefficient on complex GNNs with large-scale intermediate data 
\cite{GAT_ICLR_2018,GGCN_ICLR_2016,GGNN_ICLR_2016}, as swapping large-scale intermediate data significantly increases host-GPU communication. Moreover, since the intermediate data are swapped at a whole-graph granularity, ROC's approach may fail if a single intermediate tensor is excessively large.


\begin{table}[h]
    \vspace{-0.05in}
	\caption{Neighbor replication factor $\alpha$ under different partitions.}
	\vspace{-0.1in}
	\label{tab:dup_factor}
	\centering
	\footnotesize
	{\renewcommand{\arraystretch}{1.0}
	\begin{tabular}{c|l l l l l l l l l}
		\hline
		
		\hline
		\multirow{1}*{Partitions}
		&{\textbf{2}} &
		{\textbf{4}}  &
		{\textbf{8}}&
        {\textbf{16}}&
		{\textbf{32}}&
		{\textbf{64}}&
		{\textbf{128}}&
		{\textbf{256}}&
		{\textbf{512}}\\
		\hline
        \hline
		it-2004 &1.23&1.35&1.46&1.52&1.60&1.63&1.71&1.76&1.85\\
        ogbn-paper&1.25&1.52&2.13&3.02&4.46&6.34&8.50&10.6&12.3\\
        friendster &1.32&1.77&2.68&3.86&5.48&7.70&10.70&14.4&18.1\\
        \hline
        
        \hline
	\end{tabular}
 }\vspace{-0.06in}
\end{table}

\noindent{\textbf{{Limitation 2:}}} 
Existing systems suffer from increased host-GPU communication caused by neighbor replication. When a graph is partitioned into multiple subsets, vertices with multiple outgoing edges are replicated across partitions to serve as incoming neighbors for remote neighbor aggregation. These vertices, which we refer to as duplicate neighbors, need to be transferred individually among partitions during computation, leading to increased host-GPU communication. The communication volume is quantified by the neighbor replication factor $\alpha$, which is defined as the average number of replicas per vertex. Compared to the ideal case where each vertex is transferred only once, transferring neighbor data for each chunk individually results in $\alpha$ times communication volume, which makes the PCIe-based host-GPU communication a critical performance bottleneck. Table \ref{tab:dup_factor} presents the replication factor of the three large graphs used in our evaluation, split from 2 to 512 partitions. We can observe that increasing the number of partitions leads to an increase in data transfer volume. However, existing partition-based systems do not account for this factor. NeuGraph and NeutronStar \cite{NEUGRAPH_ATC_2019,NEUTRONSTAR_SIGMOD_2022} use host-side filters to remove unnecessary data before communicating a partition, but still need to transfer the neighbor data entirely for each of them.

\section{The \sysname framework}
\label{sec:frame}
\begin{figure}
  \vspace{-0.01in}
	\centering
	\includegraphics[width=2.8in]{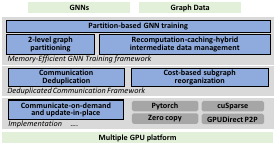}
  \vspace{-0.05in}
	\caption{\sysname system overview.
    	}
	\label{fig:sec3:overview}
 \vspace{-0.1in}
\end{figure}

\begin{figure*}[t]
	\centering
	\includegraphics[width=\textwidth]{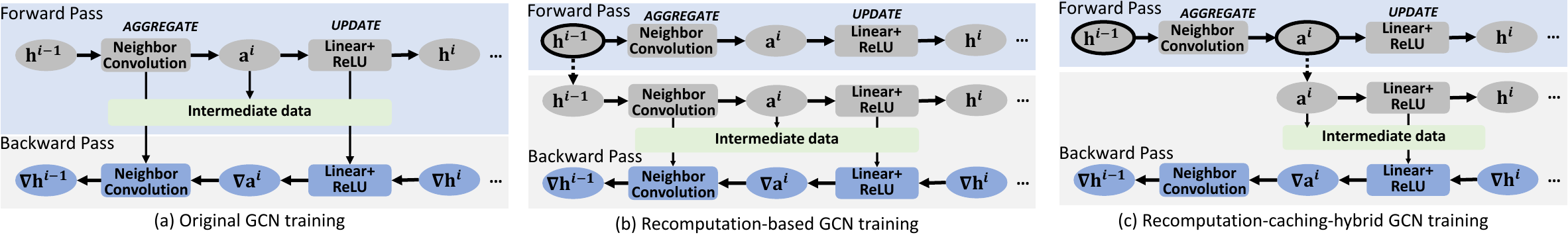}
  \vspace{-0.15in}
	\caption{{A Graphical illustration of the original, recomputation-based, and recomputation-caching-hybrid training methods on the GCN model. Here we give the example of a single layer, while other layers have the same calculation mode.} A box represents an operation, a circle represents a tensor, and a tensor surrounded by a black frame indicates that the it needs to be cached in the CPU memory (checkpoint). Solid arrows indicate dependencies between tensors and operations and dash arrows indicate host-GPU communication.}
 \label{fig:sec3:rc_gcn}
 \vspace{-0.1in}
\end{figure*}


We present \sysname, a GPU-accelerated full-graph GNN system that addresses the limitations outlined in Section \ref{sec2:frame} through two critical system components. First, \sysname provides a memory-efficient training framework that reduces the memory consumption of both vertex data and intermediate data. Second, \sysname provides a deduplicated communication framework that effectively reduces host-GPU communication for duplicated neighbor access among subgraphs.
Figure \ref{fig:sec3:overview} provides an architectural overview of \sysname. 

\Paragraph{Memory-efficient GNN training framework} \revision{\sysname adopts a graph partitioning method that groups edges incident on the same destination into a single chunk. This design facilitates full neighbor aggregation on each chunk individually, enabling \sysname to support complex GNNs (such as GAT \cite{GAT_ICLR_2018} and GGCN \cite{GGCN_ICLR_2016}) efficiently while reducing memory usage. 
} Moreover, \sysname extends the recomputation-based DNN training method \cite{SUBMEM_ARXIV_2016} to GNN training, which avoids storing intermediate data by recomputing it in the backward pass. Taking the advantages that some graph operations involve only simple edge computation and do not generate intermediate data, we hybrid GPU-based recomputation and CPU-based data caching to reduce the additional processing overhead.

\Paragraph{Deduplicated communication framework} We observe that the duplicated data access between concurrently scheduled subgraphs on multiple GPUs and the duplicated data access between sequentially scheduled subgraphs on the same GPU can benefit from inter-GPU and intra-GPU data communication, both of which have higher speeds compared to PCIe-based host-GPU communication. We propose a deduplicated communication method that transfers the data of each duplicated neighbor only once between CPU and GPU, 
and converts redundant host-GPU communication into more efficient inter-GPU communication or intra-GPU data reuse. Moreover, considering the impact of vertex distribution on communication deduplication effectiveness, \sysname incorporates a cost-model guided subgraph reorganization method to minimize communication overhead.

\section{Memory-Efficient GNN Training Framework}


\subsection{Edge-Cut 2-Level Graph Partitioning}

\begin{figure}
	\centering
 \vspace{-0.0in}
	\includegraphics[width=0.7\linewidth]{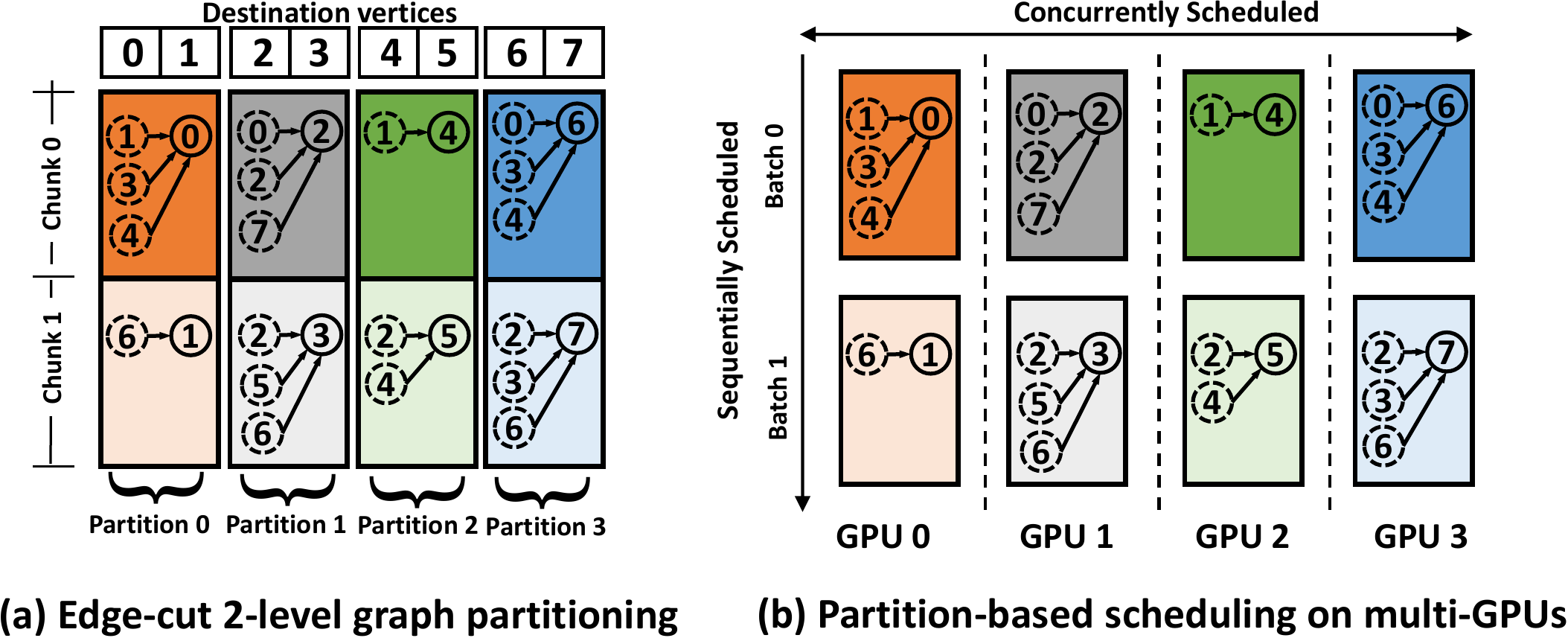}
 \vspace{-0.05in}
	\caption{An illustration of 2-level graph partitioning and the task scheduling on multiple GPUs.
    	}
     \vspace{-0.05in}
	\label{fig:sec3:twod}
\end{figure}

\label{sec3:partitioning}
\sysname employs an edge-cut partitioning to split the graph into small execution units suitable for processing by a single GPU, as shown in Figure \ref{fig:sec3:twod} (a). Initially, the input graph is split into $m$ (the number of GPUs) partitions through Metis partitioning \cite{metis_ipps_1996} to improve load balancing and group closely linked vertices into one partition. Each partition is subsequently divided into computation-balanced chunks through range-based partitioning \cite{GEMINI_OSDI_2016}, with each chunk containing a unique set of destination vertices and their associated edges. This partitioning method enables the full-neighbor aggregation to be implemented on each chunk individually. \revision{It is worth mentioning that only the in-edges of destinations need to be grouped, as the complex aggregations are executed only in the forward pass. In backward propagation, source vertices accumulate the gradient along the out-edges through summation. Leveraging the associativity of the sum operation, multiple source replicas in different chunks can independently calculate gradients and subsequently aggregate them.} During GNN training, partitioned subgraphs are scheduled in a fixed order as shown in Figure \ref{fig:sec3:twod} (b). Chunks belong to the same partition are sequentially scheduled on one GPU, and chunks with the same local position from different partitions are scheduled concurrently on different GPUs. For brevity, we use the term \emph{batch} to refer to a group of concurrently scheduled chunks from different partitions.

\subsection{Recomputation-Caching-Hybrid Intermediate Data Management}
\label{sec3:rc}
To reduce the memory consumption of DNN training, researchers have proposed a recomputation-based strategy that eliminates the need to store intermediate data for every layer by recalculating an additional forward pass in the backward computation \cite{SUBMEM_ARXIV_2016}. However, this method is designed for DNN training and assumes the training data of all layers can be entirely stored in GPU memory as the checkpoint. This makes it unsuitable for full-graph GNN training, where the training data of the entire graph can occupy a significant amount of memory. \revision{We generalize the recomputation-based approach to the CPU-GPU heterogeneous platform. Figure \ref{fig:sec3:rc_gcn} (b) shows a graphical illustration of the recomputation-based method on a single layer of GCN training, where the \textbf{AGGREGATE} operation is the neighbor convolution and the \textbf{UPDATE} operation is the \texttt{Linear+ReLU} calculation.} In the forward computation of each GNN layer, \sysname copies the output representations to CPU memory as checkpoint and releases the intermediate data to make room for training the next \emph{batch}. In the backward pass, \sysname loads the checkpoint from CPU, recomputes the forward pass, and computes the gradients based on the regenerated intermediate data. This method allows \sysname to store the training data of only one layer, thereby reducing the overall GPU memory consumption. Importantly, the recomputation-based approach maintains the accuracy of the original training method \cite{SUBMEM_ARXIV_2016} as shown in Figure \ref{fig:sec3:rc_gcn} (a), because the regenerated intermediate data are identical to that produced in the forward computation.

\revision{Recomputation-based training reduces memory consumption but entails an additional forward pass. However, not all recomputation is necessary. In the case of GNNs with simple arithmetic edge computation, where the \textbf{AGGREGATE} operation does not yield intermediate results required for gradient computation, caching the output of the \textbf{AGGREGATE} operation in CPU can eliminate the need for recomputation. For instance, the GCN model \cite{GCN_ICLR_2017} in Equation \ref{eq:gcn} employs a weighted neighbor summation as the \textbf{AGGREGATE} operation. Recomputing the \textbf{AGGREGATE} requires loading representations of all neighbors from the CPU and redoing neighbor convolution on GPUs, resulting in $O(\alpha |V|)$ CPU-GPU communication and $O(|E|)$ GPU computation. Alternatively, caching the output neighbor representations of \textbf{AGGREGATE} in CPU memory and transferring them back when needed achieves the same functionality with only $O(|V|)$ host-GPU communication. Based on this observation, we propose a recomputation-caching-hybrid method shown in Figure \ref{fig:sec3:rc_gcn} (c). In the forward pass, \sysname caches the neighbor representation ($\textbf{a}^i$) in the CPU as the recomputation checkpoint. In the backward pass, \sysname skips the \textbf{AGGREGATE} step, loads the cached neighbor representations from the CPU, and recomputes only the \textbf{UPDATE} stage.}
This hybrid design can benefit a broad range of popularly used GNNs, such as GCN \cite{GCN_ICLR_2017}, GraphSage \cite{GRAPHSAGE_NIPS_2017}, GIN \cite{GIN_ICLR_2019}, and CommNet \cite{COMMNET_NIPS_2016}. However, for GNNs with neural network computation on edges (e.g., GAT \cite{GAT_ICLR_2018} and GGCN \cite{GGCN_ICLR_2016}), the overhead of caching the $O(|E|)$ intermediate data can be higher than that of recomputation. In such cases, \sysname falls back to the recomputation-based method as depicted in Figure \ref{fig:sec3:rc_gcn} (b).

\begin{algorithm}[t]\small
	\caption{Workflow of \sysname for a single epoch}  
	\label{alg:exec_flow}  
	\begin{algorithmic}[1]
		\Require Graph $G(V,E)$, feature $\{\textbf{h}^{0}_v\mid v\in V\}$,
		Initial parameterized GNN layers $\{\textbf{GNN}^0, \textbf{GNN}^1 \ldots \textbf{GNN}^{L-1}\}$
		\Ensure Updated parameterized GNN layers $\{\textbf{GNN}^0, \textbf{GNN}^1 \ldots \textbf{GNN}^{L-1}\}$
        \State $\{G_{ij}$$\mid$$ 0$$\leq$$i$$<$$m,0$$\leq$$j$$<$$n\}$=\textbf{\texttt{two\_level\_partition}}($G$, $m$, $n$)
  
		\State $\{\mathcal{G}_{ij}$$\mid$$ 0$$\leq$$i$$<$$m,0$$\leq$$j$$<$$n\}$= \textbf{\texttt{deduplication}}($\{G_{ij}$$\mid$$ 0$$\leq$$i$$<$$m,0$$\leq$$j$$<$$n\}$)
		\State allocate $\{\textbf{h}^{l}, \nabla \textbf{h}^{l}\mid 0\leq l\leq L\}$ in the CPU memory.
  
		\For{layer $l$ = $0$ to $L-1$}
		    \For{\emph{batch} with id $j$ = $0$ to $n-1$}
                \State 
        $\{\textbf{h}^{l}_{N_{ij}}$$\mid$$ 0$$\leq$$i$$<$$m\}$ $\leftarrow$
\textbf{\texttt{dedup\_comm\_fwd}}{($\textbf{h}^{l}$, $\{{\mathcal{G}}_{ij}$$\mid$$ 0$$\leq$$i$$<$$m\}$, \texttt{HtoD})}
        
        \For{\texttt{GPU} $i=0$ to $m-1$} \textbf{in parallel}
		\State {$\textbf{h}^{l+1}_{V_{ij}}$ =\enspace \texttt{GPU(}$i$\texttt{).forward}($\textbf{GNN}^i$,$\textbf{h}^{l}_{N_{ij}}$, $\mathcal{G}_{ij}$)}
		\State 
    {$\textbf{h}^{l+1}$$\leftarrow$ \texttt{GPU(}$i$\texttt{).mem\_copy}($\textbf{h}^{l+1}_{V_{ij}}$, $V_{ij}$, \texttt{DtoH})} 
	\EndFor
        \EndFor
		  \EndFor
    \vspace{-0.03in}
		\State $\textbf{loss}$= \texttt{downstream\_task}($\textbf{h}^{L}$)
		\State $\nabla \textbf{h}^{L}$ =$\textbf{loss}$\texttt{.backward()} 
		\For{layer $l$ = $L-1$ to $0$ }
		    \For{\emph{batch} with id $j$ = $0$ to $n-1$}
      \State 
        $\{\textbf{chkpt}^{l}_{ij}$$\mid$$ 0$$\leq$$i$$<$$m\}$ $\leftarrow$\textbf{\texttt{load\_recomp\_chkpt}}{($\ldots$, \texttt{HtoD})}
      \For{GPU $i=0, 1, \ldots,m-1$} \textbf{in parallel}
            \State 
            {$\nabla \textbf{h}^{l+1}_{V_{ij}}$$\leftarrow$ \texttt{GPU(}$i$\texttt{).mem\_copy}($\nabla\textbf{h}^{l+1}$, $V_{ij}$, \texttt{HtoD})}
		    \State 
            {\quad\quad\quad\quad\texttt{GPU(}$i$\texttt{).reforward}($\textbf{GNN}^{l}$, $\textbf{chkpt}^{l}_{ij}$ $\mathcal{G}_{ij}$)} 
		    \State 
            {$\nabla \textbf{h}^{l}_{N_{ij}}$= \enspace\texttt{GPU(}$i$\texttt{).backward}($\textbf{GNN}^{l}$, $\nabla \textbf{h}^{l+1}_{V_{ij}}$, $\mathcal{G}_{ij}$)}
		\EndFor
  	\State 
            {$\nabla \textbf{h}^{l}$$\overset{\oplus}{\leftarrow}$\textbf{\texttt{dedup\_comm\_bwd}}($\{\nabla\textbf{h}^{l}_{N_{ij}}$, $\mathcal{G}_{ij}$$\mid$$ 0$$\leq$$i$$<$$m\}$,\texttt{DtoH})}
        \EndFor
		\EndFor
  \vspace{-0.03in}
    \For{layer $l$ = $0$ to $L-1$}
        \State \texttt{sync\_and\_update} ($\textbf{GNN}^l$) \textcolor{gray}{//parameter update}
    \EndFor
	\end{algorithmic} 
\end{algorithm}

\subsection{Overall Execution Flow in \sysname}
Algorithm \ref{alg:exec_flow} outlines the overall execution flow. To begin with, \sysname partitions the graph with 2-level partitioning (line 1). Each subgraph, represented by $G_{ij}$, consists of a set of disjointly split vertices $V_{ij}$ and their incoming edges $E_{ij}$, where $i$ is the partition id and $j$ is the chunk id. After graph partitioning, the communication deduplication module reorganizes the partitioned subgraphs, deduplicates the neighbor accesses, and generates the new partitions $\{\mathcal{G}_{ij}$$\mid$$ 0$$\leq$$i$$<$$m,0$$\leq$$j$$<$$n\}$ for parallel training (line 2). Since the initial partitioned graph $G_{ij}$ is no longer used in the subsequent computation, we use $V_{ij}$ and $E_{ij}$ to represent the vertices and edges of the new subgraph $\mathcal{G}_{ij}$ and denotes its in-neighbor set by $N_{ij}$. After preprocessing, \sysname initializes the vertex representation buffer $\textbf{h}^l$ and gradient buffer $\nabla \textbf{h}^l$ in the CPU memory (line 3).

In the training process, \emph{batches} are scheduled sequentially, and subgraphs in each \emph{batch} are processed in parallel.
In the forward pass of each \emph{batch}, neighbor representations of all subgraphs, i.e., $\{\textbf{h}^{l}_{N_{ij}} $$\mid$$0$$\leq$$i$$\le$$m\}$ are first loaded from CPU to GPUs through the deduplicated communication framework (line 6) which will be discussed in Section \ref{sec3:repeat}. Following this, each GPU performs the forward computation (lines 7-8), copies the newly computed vertex representation $\textbf{h}^{l+1}_{{V}_{ij}}$ to CPU (line 9), and releases intermediate data to make room for training the next \emph{batch}. After completing the forward pass, the downstream task takes the final layer output $\textbf{h}^{L}$ as input, computes the \texttt{loss} and the gradient of \texttt{loss} to the final layer representation, i.e., $\nabla\textbf{h}^L$ (lines 10-11). In the backward pass, computation is scheduled from the last layer to the first layer (line 12). In each \emph{batch}, \sysname reloads the checkpoint to GPUs (line 14), loads the gradients of destinations from CPU memory (line 16), recomputes the forward pass of the current layer (line 17), and computes the gradients (line 18) based on the regenerated intermediate data. Finally, the neighbor gradients $\{\nabla \textbf{h}^l_{N_{ij}}\mid 0$$\leq$$i$$\le$$m\}$ are transferred back to CPU and accumulated into the gradient buffer $\nabla\textbf{h}^l$ through deduplicated communication (line 19).

After completing a forward-backward pass, \sysname updates the model parameters according to the gradients (line 19). Since the model parameters in GNN are often small, \sysname replicates them among GPUs and uses \texttt{all\_reduce()} function to synchronize.

\begin{figure*}[t]
  \vspace{-0.02in}
  	\centering
	\includegraphics[width=\textwidth]
 {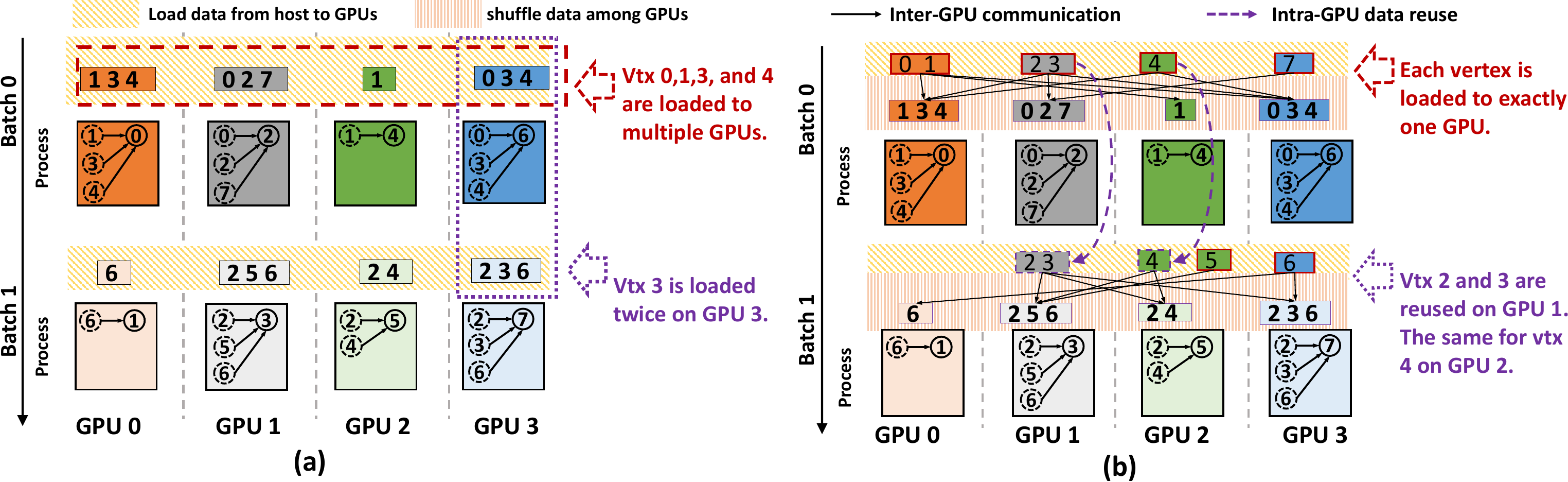}
 \vspace{-0.08in}
	\caption{Graphical illustrations of (a) increased host-GPU communication caused by duplicated neighbor accesses and (b) the proposed communication deduplication method. 
    	}
     	\label{fig:sec3:graph_partitioning_dup}
     \vspace{-0.08in}
\end{figure*}

\Paragraph{The Effectiveness of memory reduction}
 By combining 2-level graph partitioning and recomputation-based training, \sysname can maintain the training data of one GNN layer for a single subgraph in each GPU. In an ideal case where the graph is evenly partitioned, the vertex data volume of one subgraph can be formalized as $(1+\alpha_{m*n})|V|/(m*n)$, where $m*n$ is the number of subgraphs, $|V|$ is the number of vertices, and $\alpha_{(m*n)}$ is the neighbor replication factor given $m*n$. As shown in Table \ref{tab:dup_factor}, every doubling of the number of partitions results in a 47\%, 35\%, and 32\% reduction in the memory consumption of vertex data for the three graphs, respectively. The memory consumption of intermediate data varies, depending on the GNN model in use. Some models are dominated by the number of vertices \cite{GCN_ICLR_2017,GRAPHSAGE_NIPS_2017,COMMNET_NIPS_2016,GIN_ICLR_2019}, while others are dominated by the number of edges \cite{GGCN_ICLR_2016,GGNN_ICLR_2016,GAT_ICLR_2018}, and both decrease linearly as $m*n$ increases. In practical training, memory consumption can be adjusted by tuning the number of partitions to adapt to different GPUs.






\section{Deduplicated Communication Framework}
\label{sec3:repeat}

In this section, we present the design and implementation of deduplicated communication framework. 

\subsection{Basic Design}
\Paragraph{Inter-GPU duplicated neighbor access}
Duplicated neighbors between concurrently-scheduled subgraphs cause the same vertex to be transferred to multiple GPUs. As indicated by the red dashed box in Figure \ref{fig:sec3:graph_partitioning_dup} (a), vertex 0, 1, 3, and 4 are transferred to multiple GPUs in \emph{batch} 0. The data of these vertices are redundantly communicated between CPU and GPUs. Instead, we can transfer the duplicated vertex to one GPU and handle the access requests from other GPUs through inter-GPU communication. Benefiting from the high communication bandwidth between GPUs (as described in Section \ref{sec:2}), converting host-GPU communication to inter-GPU communication can significantly improve performance. 


\Paragraph{Intra-GPU duplicated neighbor access} 
Duplicated neighbors between sequentially-scheduled subgraphs cause the same vertex to be transferred multiple times to the same GPU. As indicated by the purple dotted boxes in Figure \ref{fig:sec3:graph_partitioning_dup} (a), vertex 2 and 5 in GPU 1 and vertex 3 in GPU 3 are loaded in both batch 0 and batch 1. For those adjacently-scheduled subgraphs, neighbor access to duplicated neighbors from the successor subgraph can directly reuse the already transferred data in GPU, converting host-GPU communication to intra-GPU data access. 

\Paragraph{Communication Deduplication} 
We stack these two techniques to cooperatively reduce the communication for duplicated neighbor accesses, as illustrated in Figure \ref{fig:sec3:graph_partitioning_dup} (b). Our method involves two steps. In the first step, it computes the union of neighbors in each \emph{batch} (i.e., a group of concurrently scheduled subgraphs). This union is then deduplicated and stored in a transition vertex set denoted by $\mathbb{N}^{\cup}_{j}=\cup_{i=0}^{m}N_{ij}$, where $0$$\leq$$j$$<$$m$ indicates the batch id. During computation, each vertex in the transition vertex set is transferred to exactly one GPU and shared among GPUs through inter-GPU communication. Figure \ref{fig:sec3:graph_partitioning_dup} (b) provides a graphical example. The deduplicated vertex sets, $\mathbb{N}^{\cup}_{0}$, i.e., $\{0,1,2,3,4,7\}$, and $\mathbb{N}^{\cup}_{1}$, i.e., $\{2,3,4,5,6\}$ are transferred only once, reducing host-GPU communication times from 19 to 11. To share communication workload among GPUs, we divide $\mathbb{N}^{\cup}_{j}$ into $m$ subsets $\{\mathbb{N}_{0j},\ldots \mathbb{N}_{m-1j}\}$, and assign the communication of $\mathbb{N}_{ij}$ to GPU $i$, where $\mathbb{N}_{ij}$ is the subset of $\mathbb{N}^{\cup}_{j}$ belonging to partition $i$, as shown in Figure \ref{fig:sec3:graph_partitioning_dup} (b). In the second step, we perform the intra-GPU deduplication on the transition vertex set for each pair of adjacently-scheduled subgraphs, e.g., $\mathbb{N}_{ij-1}$ and $\mathbb{N}_{ij}$. We divide the successor transition vertex set $\mathbb{N}_{ij}$ into two disjoint subsets $\mathbb{N}^{gpu}_{ij}$ and $\mathbb{N}^{cpu}_{ij}$, where $\mathbb{N}^{gpu}_{ij}$ represents the duplicated vertices, i.e., $\mathbb{N}_{ij}\cap \mathbb{N}_{ij-1}$, and $\mathbb{N}^{cpu}_{ij}$ represents the remaining vertices $\mathbb{N}_{ij}\setminus \mathbb{N}_{ij-1}$. When loading the data of $\mathbb{N}_{ij}$ from the CPU, vertices in $\mathbb{N}^{gpu}_{ij}$ are directly reused from the GPU, and vertices in $\mathbb{N}^{cpu}_{ij}$ are loaded from the CPU memory. Figure \ref{fig:sec3:graph_partitioning_dup} (b) shows a graphical illustration with purple dashed arrows. When processing batch 1, the data of $\mathbb{N}^{gpu}_{11}$ ($\{2, 3\}$) and $\mathbb{N}^{gpu}_{12}$ ($\{4\}$) are directly reused from \emph{batch} 0, and the data of $\mathbb{N}^{cpu}_{12}$ ($\{5\}$) and $\mathbb{N}^{cpu}_{13}$ ($\{6\}$) are loaded from CPU memory. This step further reduces host-GPU communication times from 11 to 8.

\subsection{Workflow of Deduplicated Communication} 

\sysname uses a transition data buffer $\textbf{h}_{\mathbb{N}_{ij}}$ on each GPU to manage the data of transition vertices $\mathbb{N}_{ij}$, based on which we can decouple host-GPU communication and inter-GPU communication.

{
\begin{algorithm} \footnotesize
	\caption{\texttt{dedup\_comm\_fwd}}
	\label{alg:comm:fwd}  
	\begin{algorithmic}[1] 
		\Require 
  $\{N_{ij}, \mathbb{N}_{ij}, \mathbb{N}^{gpu}_{ij}, \mathbb{N}^{cpu}_{ij} \mid 0$$\leq$$i$$<$$m$$\}$, $\textbf{h}^{l}$ in the CPU.
  
		\Ensure load $\{\textbf{h}_{N_{ij}}$ $\mid 0$$\leq$$i$$<$$m\}$ to the $m$ GPUs separately
		\For{\texttt{GPU} $i=0$ to $m-1$} \textbf{in parallel} \textcolor{gray}{//host-to-GPU}
        \State $\textbf{h}_{\mathbb{N}_{ij}}\leftarrow$ \texttt{GPU(}$i$\texttt{).reuse(}$\textbf{h}_{\mathbb{N}_{ij-1}}$,$\mathbb{N}^{gpu}_{ij}$)
        \State $\textbf{h}^l_{\mathbb{N}_{ij}}\leftarrow$ \texttt{GPU(}$i$\texttt{).mem\_copy\_sparse(}$\textbf{h}^{l}$,$\mathbb{N}^{cpu}_{ij}$, \texttt{HtoD})
        \EndFor
        \vspace{-0.05in}
        \State \texttt{synchronize()}
        \For{\texttt{GPU} $i=0$ to $m-1$} \textbf{in parallel} \textcolor{gray}{//GPU-to-GPU}
        \For{\texttt{GPU} $k=i+1$ to $(m+i)$ \texttt{mod} $m$}
        \State $\textbf{h}_{N_{ij}}\leftarrow$ \texttt{GPU(}$i$\texttt{).fetch\_from\_gpu}($k$, $\textbf{h}_{\mathbb{N}_{ik}}$, ${N}_{ij}\cap \mathbb{N}_{ik}$, \texttt{DtoD})
        \EndFor
        \EndFor
        \vspace{-0.03in}
        \State \texttt{synchronize()}
	\end{algorithmic} 
\end{algorithm}
}


Algorithm \ref{alg:comm:fwd} outlines the workflow of deduplicated communication in the forward pass. It loads the neighbor representations $\textbf{h}^l_{N_{ij}}$ from the CPU data buffer $\textbf{h}^l$ to the $m$ GPUs separately, based on four distinct vertex sets: $\{ N_{ij}, \mathbb{N}_{ij}, \mathbb{N}^{gpu}_{ij}, \mathbb{N}^{cpu}_{ij}$$\mid$$0$$\leq$$i$$<$$m$$\}$. In the first step, each GPU $i$ loads the data of transition vertices to the transition data buffer $\textbf{h}_{\mathbb{N}_{ij}}$, by reusing the data of $\mathbb{N}^{gpu}_{ij}$ from $\textbf{h}_{\mathbb{N}_{ij-1}}$ and loading the data of $\mathbb{N}^{cpu}_{ij}$ from the CPU (lines 2-3). In the second step, GPUs communicate with each other to fetch the data of each $N_{ij}\cap\mathbb{N}_{ik}$ from remote transition data buffers and assemble the neighbor data $\textbf{h}^l_{N_{ij}}$ in local memory (lines 5-7).


{
\begin{algorithm} \footnotesize
	\caption{\texttt{dedup\_comm\_bwd}}
	\label{alg:comm:bwd}  
	\begin{algorithmic}[1] 
		\Require  $\{ N_{ij}, \mathbb{N}_{ij},\mathbb{N}^{gpu}_{ij}, \mathbb{N}^{cpu}_{ij}$$\mid$$0$$\leq$$i$$<$$m$$\}$, $\{\nabla\textbf{h}^l_{N_{ij}}$$\mid$$0$$\leq$$i$$<$$m$$\}$ on the $m$ GPUs
		\Ensure Accumulate $\{\textbf{h}_{N_{ij}}$ $\mid$$ 0$$\leq$$i$$<$$m\}$ to the CPU gradient buffer $\nabla\textbf{h}^{l}$.
        \For{\texttt{GPU} $i=0$ to $m-1$} \textbf{in parallel}
        \For{\texttt{GPU} $k=i+1$ to $(m+i)$ \texttt{mod} $m$}
        \State $\nabla\textbf{h}_{\mathbb{N}_{ik}}\overset{\oplus}{\leftarrow}$ \texttt{GPU(}$i$\texttt{).accum\_to\_gpu}($k$, $\nabla\textbf{h}^l_{N_{ij}}$, ${N}_{ij}\cap \mathbb{N}_{ik}$, \texttt{DtoD})
        \EndFor
        \EndFor
        \vspace{-0.03in}
        \State \texttt{synchronize()}
		\For{\texttt{GPU} $i=0$ to $m-1$} \textbf{in parallel}
        \State $\nabla\hat{\textbf{h}}_{\mathbb{N}^{cpu}_{ij}}\leftarrow$ \texttt{GPU(}$i$\texttt{).mem\_copy\_sparse(}$\nabla \textbf{h}_{\mathbb{N}_{ij}}$, $\mathbb{N}^{cpu}_{ij}$, \texttt{DtoH})
        \State $\nabla\textbf{h}^{l}\overset{\oplus}{\leftarrow}\nabla\hat{\textbf{h}}_{\mathbb{N}^{cpu}_{ij}}$ \textcolor{gray}{//CPU computation}
        \EndFor
        \vspace{-0.05in}
        \State \texttt{synchronize()}
	\end{algorithmic} 
\end{algorithm}
}

In the backward pass, deduplicated communication function copies the neighbor gradients $\nabla\textbf{h}^l_{N_{ij}}$ back to the CPU and accumulates them to the gradient buffer $\nabla \textbf{h}^l$. Algorithm \ref{alg:comm:bwd} outlines this process. In the first step, GPUs accumulate the neighbor gradients $\nabla\textbf{h}_{N_{ij}}$ back to the gradient buffer of transition vertices (lines 1-3). Subsequently, each GPU moves the gradients of $\mathbb{N}^{cpu}_{ij}$ out to CPU memory (line 6) and reserves the gradient of $\mathbb{N}^{gpu}_{ij}$ in GPU to accumulate the gradients of the next \emph{batch}. One the CPU side, the gradients of $\mathbb{N}^{cpu}_{ij}$ are accumulated to $\nabla \textbf{h}^l_i$ with CPUs (line 7). Since the gradient accumulation only involves simple arithmetic addition, utilizing CPUs is faster than copying the data to the GPU computation, in which the data movement involves bidirectional host-GPU communication.


\vspace{-0.07in}
\subsection{Cost-Effective Subgraph Reorganization}
\label{sec:5:cost}
The effectiveness of communication deduplication is affected by the distribution of duplicated neighbors. To enhance communication efficiency, we quantify the deduplicated communication overhead and propose a subgraph reorganization method to minimize it.


\Paragraph{Cost of deduplicated communication} Initially, the neighbor set of each subgraph is transferred entirely. The host-GPU communication has a volume of $\textbf{V}_{ori}=\sum_{j=0}^{n}\sum_{i=0}^{m}|N_{ij}|$, where $|N_{ij}|$ represents the number of neighbors of partition $i$ chunk $j$. By utilizing inter-GPU communication duplication, the host-GPU communication volume is reduced to $\textbf{V}_{+p2p}=\sum_{j=0}^{n}|\cup_{i=0}^{m}N_{ij}|$, where $\cup_{i=0}^{m}N_{ij}$ is the transition vertex set of batch $j$. By further applying intra-GPU communication duplication, the duplicated transition vertices of each pair of adjacent subgraphs, i.e., $\cup_{i=0}^{m}N_{ij}\cap \cup_{i=0}^{m}N_{ij-1}$, are no longer required to be transferred. Consequently, the host-GPU communication is further reduced to $\textbf{V}_{+ru}=|\cup_{i=0}^{m}N_{i0}|+\sum_{j=1}^{n}|\cup_{i=0}^{m}N_{ij}\setminus \cup_{i=0}^{m}N_{ij-1}|$. Finally, the total communication overhead can be formalized as 

\vspace{-0.1in}
\begin{align}\small
\label{eq:trans_cost}
\textbf{C}=\textbf{V}_{+ru}/\textbf{T}_{hd}+(\textbf{V}_{ori} - \textbf{V}_{+p2p})/\textbf{T}_{dd}+(\textbf{V}_{+p2p}-\textbf{V}_{+ru})/\textbf{T}_{ru},
\end{align}
where $\textbf{T}_{hd}$, $\textbf{T}_{dd}$, and $\textbf{T}_{ru}$ represent the throughput of host-GPU communication, inter-GPU communication, and intra-GPU data reusing, respectively. These parameters are environment-specific and depend on the used GPU platform.

We can observe that the communication cost $\textbf{C}$ is affected by the number of duplicated neighbors among subgraphs. Obtaining the minimal $\textbf{C}$ requires careful adjustments of vertex distribution in the partitions. However, optimizing this goal in the partitioning stage is challenging because it involves a vast search space at the vertex granularity and couples with several constraints, such as load balancing and communication reduction. To simplify this problem, we propose a subgraph granularity optimization approach. Specifically, given an initialized load-balancing optimized graph partition $\{G_{ij}\mid 0$$\leq$$i$$<$$m,0$$\leq$$j$$<$$n\}$, the objective is to find a reorganized partition $\{\mathcal{G}_{ij}\mid 0$$\leq$$i$$<$$m,0$$\leq$$j$$<$$n\}$ that minimizes the cost $\textbf{C}$ in Equation \ref{eq:trans_cost}, where each $\mathcal{G}_{ij}$ is a subgraph from the initial partition, e.g., ${G}_{kl}$. This combinatorial optimization problem is NP-hard as it can be reduced to a variant of the classical NP-hard traveling salesman problem \cite{TSP_BOOK_1995}, which aims to find a Hamiltonian circuit in a weighted undirected complete graph that minimizes the total weight of the circuit. Therefore, it is infeasible to obtain an optimal solution in polynomial time using an exact algorithm. Next, we propose a 2-phase heuristic to reorganize the partition.




{\begin{algorithm}[t] \footnotesize
	\caption{Partition reorganization}
	\label{alg:greedy}  
	\begin{algorithmic}[1] 
		\Require initial partitions $\{G_{ij}\mid 0$$\leq$$i$$<$$m,0$$\leq$$j$$<$$n\}$
		\Ensure reorganized partitions $\{\mathcal{G}_{ij}\mid 0$$\leq$$i$$<$$m,0$$\leq$$j$$<$$n\}$
  \newline\underline{\emph{Phase 1: Reorganization for maximizing inter-GPU duplication}}
        \For{$j=0$ to $n-1$} 
        \State $G^t_{0j}\leftarrow G_{0j}$; $\mathbb{N}^{\cup}_{j}\leftarrow N_{0j}$
        \EndFor
        \vspace{-0.05in}
        \For{$i=1$ to $m-1$}
        \State $\mathcal{K}\leftarrow \{0, 1, \ldots,n$$-$$1\}$ \textcolor{gray}{//subgraphs that have not been processed}
        \For{$j=0$ to $n-1$} 
        \State \texttt{find} $k$ \texttt{from} $\mathcal{K}$,  \texttt{s.t.,} $\forall a\in\mathcal{K}: |N_{ik}\cap \mathbb{N}^{\cup}_{j}|\geq |N_{ia}\cap \mathbb{N}^{\cup}_{j}|$     
        \State $G^t_{ij}\leftarrow G_{ik}$;
        $\mathbb{N}^{\cup}_{j}\leftarrow \mathbb{N}^{\cup}_{j}\cup N_{ik}$; $\mathcal{K} \leftarrow \mathcal{K}\setminus k$ 
        \EndFor
        \EndFor
        \vspace{-0.05in}
        \newline\underline{\emph{Phase 2: Reorganization for maximizing intra-GPU duplication}}
        \For{$i=0$ to $m-1$}
        \State $\mathcal{G}_{i0} \leftarrow G^t_{i0}$
        \EndFor
        \vspace{-0.05in}
        \State $\mathcal{K}\leftarrow \{1, 2, \ldots,n$$-$$1\}$ \textcolor{gray}{//\emph{batches} that have not been processed}
        \For{$j=1$ to $n-1$} 
        \State \texttt{find} $k$ \texttt{from} $\mathcal{K}$,  \texttt{s.t.,} $\forall a\in\mathcal{K}: |\mathbb{N}^{\cup}_{k}\cap \mathbb{N}^{\cup}_{j-1}|\geq |\mathbb{N}^{\cup}_{a}\cap \mathbb{N}^{\cup}_{j-1}|$
        \vspace{-0.02in}
        \For{$i=1$ to $m-1$}
        \State $\mathcal{G}_{ij} \leftarrow G^t_{ik}$
        \EndFor
        \vspace{-0.05in}
        \State $\mathcal{K} \leftarrow \mathcal{K}\setminus k$ 
        \EndFor
	\end{algorithmic}
\end{algorithm}
}

\Paragraph{Partition reorganization}
We propose a 2-phase, greedy-based heuristic that optimizes communication overhead by maximizing the number of inter- and intra-GPU duplicated neighbors. The goal is to fully leverage the effect of communication deduplication.  Algorithm \ref{alg:greedy} outlines the workflow of our approach. In the first phase, we reorganize subgraphs within each partition to group subgraphs with the maximum number of duplicate neighbors into the same \emph{batch}. The objective is to maximize the number of inter-GPU duplicated neighbors while preserving the locality achieved by the Metis graph partitioning. The algorithm initializes the intermediate partition $G^t$ and the transition vertex set $\mathbb{N}^{\cup}$ for every \emph{batch} with subgraphs in partition $0$ (lines 1-2), and then reorganizes other partitions in turn. Specifically, it iterates over the transition vertex set of all \emph{batches} (line 5) and retrieves for each \emph{batch} the subgraph that has the maximum number of duplicate neighbors in the currently-processed partition (line 6). The algorithm then writes the found subgraph to the corresponding \emph{batch} in $G^t$ (line 7), and updates $\mathbb{N}^{\cup}_j$ and $\mathcal{K}$ accordingly. In the second phase, we reorganize the partition at the \emph{batch} granularity to maximize the number of intra-GPU duplicated transition vertices. The algorithm initializes $\mathcal{G}$ with \emph{batch} 0 and records other \emph{batches} in $\mathcal{K}$ (lines 8-10). During execution, it iteratively searches in $\mathcal{K}$ to find the \emph{batch} that has the maximum number of duplicated transition vertices with the current \emph{batch} (line 12), and writes the found \emph{batch} to $\mathcal{G}$. After processing all batches, we obtain the communication-efficient reorganized partition $\mathcal{G}$.

{
\Paragraph{Effectiveness with various interconnects} The proposed deduplicated communication framework offers benefits to GPU servers equipped with various interconnects. As discussed in Section \ref{sec:5:cost}, The vertex data to be transferred are divided into three subsets and handled with CPU-GPU communication, inter-GPU communication, and intra-GPU data reuse. Intra-GPU reuse consistently delivers benefits as its bandwidth $T_{ru}$ is associated with the GPU memory bandwidth and often much higher than $T_{hd}$, which is associated with the GPU-CPU interconnect bandwidth (typically using PCIe). The effectiveness of inter-GPU data sharing depends on the bandwidth of inter-GPU interconnects. Fast interconnects such as NVIDIA NVLink \cite{NVLINK} and AMD Infinity Fabric \cite{AMD_FAB}, inter-GPU communication provide substantial performance improvements through high-speed inter-GPU communication. Conversely, if GPUs are interconnected via slow PCIe, resulting in $T_{hd}$ being equal to $T_{dd}$, inter-GPU communication does not bring enhancements. Nevertheless, employing the intra-GPU reuse optimization alone still yield considerable reductions in data transfer. As shown in Table 8, the intra-GPU duplication accounts for 36\%-84\% of the total duplication volume.
}

\section{Implementation}


\label{sec3:impl_comm}
The use of deduplicated communication can significantly reduce the volume of host-GPU communication, but achieving high performance requires careful implementation, particularly for irregular memory access during communication. First, transferring the data of $\mathbb{N}_{ij}$ and $N_{ij}$ among CPU and GPUs involves non-continuous memory access. Conventional communication methods, such as NCCL \cite{NCCL} and \texttt{cudaMemcpy}, are unsuitable for our task as they are designed to operate on contiguous memory. Designing additional data compaction modules can increase the CPU overhead \cite{NEUGRAPH_ATC_2019,PYTORCHDIRECT_VLDB_2021}. Second, switching data from $\textbf{h}_{\mathbb{N}_{ij-1}}$ to $\textbf{h}_{\mathbb{N}_{ij}}$ needs to reserve the data of $\mathbb{N}^{gpu}_{ij}$ from $\textbf{h}_{\mathbb{N}_{ij-1}}$ and load the data of $\mathbb{N}^{cpu}_{ij}$ from CPU to $\textbf{h}_{\mathbb{N}_{ij}}$. This process causes random memory manipulation on the two data buffers. To address these issues, \sysname provides a high performance communication implementation with two features: \textbf{communicate-on-demand} and \textbf{-update-in-place}.

\Paragraph{On-demand communication} \sysname employs zero-copy memory access \cite{EMOGI_VLDB_20} and GPUDirect P2P access \cite{WHOLEGRAPH_SC_2022}, which allow GPUs to directly access the memory of CPUs and other GPUs within the CUDA kernel by mapping them to the same memory address.  Moreover, we implement the coalesced-and-aligned memory access optimization \cite{EMOGI_VLDB_20,PYTORCHDIRECT_VLDB_2021}, which optimize the PCIe bandwidth utilization by enabling each warp of threads to access the contiguous dimension of data. In this way, irregular and non-continous data communication among CPU and GPUs can be performed efficiently.

\begin{figure}[t]
	\centering  
 \vspace{-0.05in}
	\includegraphics[width=0.650\linewidth]{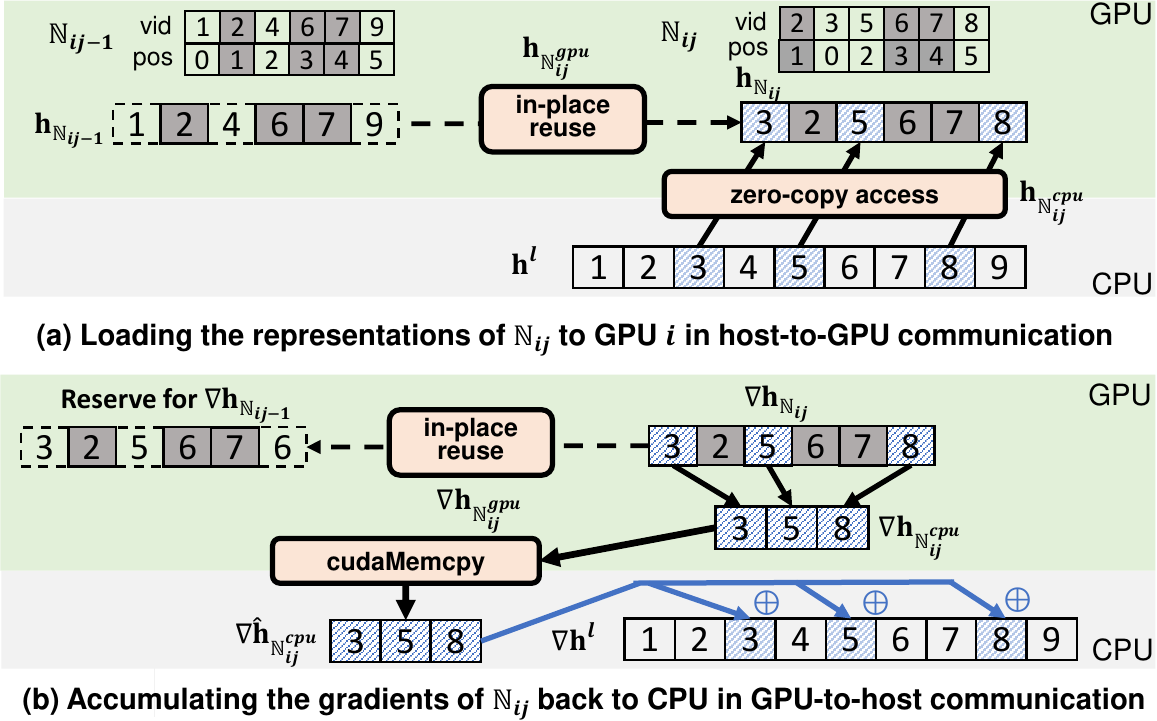}
     \vspace{-0.08in}
	\caption{Implementation of host-GPU communication.}
	\label{fig:sec3:comm_impl}
 \vspace{-0.05in}
\end{figure}

\Paragraph{In-place transition data management} \sysname uses a single data buffer to maintain the transition vertex data for all subgraphs in a partition, and $n$ position indices for maintaining the write position of transition vertices $\{\mathbb{N}_{ij}$$\mid$$0$$\leq$$j$$<$$n\}$ in the buffer.  When scheduling a new \emph{batch}, the data of newly scheduled transition vertices ($\mathbb{N}_{ij}$) are write to the buffer according to the indices. In the preprocessing, we process the transition indices for all subgraphs, making the duplicated vertices of each pair of adjacently-scheduled subgraphs have the same write positions. This allows the data of $\mathbb{N}^{gpu}_{ij}$ to be reused in-place. The data of $\mathbb{N}^{cpu}_{ij}$, which are loaded from the CPU, are inserted into the buffer based on their write positions in the indices. Figure \ref{fig:sec3:comm_impl} (a) shows a example of data loading in the host-to-GPU communication. Duplicated vertices between ${\mathbb{N}_{ij-1}}$ and ${\mathbb{N}_{ij}}$, i.e., $\{2, 6, 7\}$ have the same positions in the transition data buffer. When updating $\textbf{h}_{\mathbb{N}_{ij-1}}$ to $\textbf{h}_{\mathbb{N}_{ij}}$, the data of these vertices are reused in-place. In contrast, the data loaded from CPU, $\{3,5,8\}$, are inserted into the positions of discarded vertices $\{1,4,9\}$. 

{
\Paragraph{In-place neighbor data management} \sysname uses a single data buffer to maintain the neighbor data for all subgraphs in a partition. It uses a neighbor index for each neighbor set $N_{ij}$ to track their read positions in the local/remote transition data buffer and the write positions in the local neighbor data buffer. When switching the neighbor data between subgraphs, data of $N_{ij}$ are exchanged between the transition data buffer and neighbor data buffer according to the indices. We notice that duplicated neighbors between each pair of adjacently-scheduled subgraphs (i.e., $N_{ij}\cap N_{ij-1}$) are redundantly transferred. To address this issue, we extend the data reuse technique to inter-GPU communication. \sysname reorders the neighbor vertices of all subgraphs, making the duplicated vertices ($N_{ij}\cap N_{ij-1}$) have the same positions in the neighbor data buffer, and reuses the data of them during communication.
}

\Paragraph{Host-GPU communication implementation}
\sysname allocates both $\textbf{h}^l$ and $\nabla\textbf{h}^l$ on pinned memory with \texttt{cudaMallocHost()} to support zero-copy memory access in the host-to-GPU communication. In the GPU-to-host communication, \sysname accumulates the gradients of $\mathbb{N}^{cpu}_{ij}$ to $\nabla \textbf{h}^l$ in CPU and reserves the gradients of $\mathbb{N}^{gpu}_{ij}$ in GPU. To leverage GPU's high memory bandwidth, \sysname implements a GPU-based compaction module as shown in Figure \ref{fig:sec3:comm_impl} (b). The gradients to be moved out are first collected in GPU memory and then transferred back to CPU using \texttt{cudaMemcpyAsync()}. 

\Paragraph{Inter-GPU communication implementation}
\sysname enables GPUDirect P2P access through the \texttt{cudaDeviceEnablePeerAccess()} function, which facilitates direct memory access between GPUs. In the forward pass, \sysname uses pull-based communication, where each vertex in $N_{ij}$ reads its representation from the corresponding GPU. In the backward pass, \sysname employs a push-based communication scheme to accumulate the gradients of $N_{ij}$ back to the transition data buffer in the corresponding GPUs, utilizing the \texttt{atomicAdd\_system()} function \cite{NVLINK}. To avoid resource contention caused by multiple GPUs accessing the data from the same GPU, we implement interleaved communication optimization \cite{GEMINI_OSDI_2016} that avoids different GPUs accessing one GPU at the same time slot, as shown in Algorithm \ref{alg:comm:fwd} (line 6) and Algorithm \ref{alg:comm:bwd} (line 2).

\Paragraph{Data buffer deduplication} Maintaining the data buffer of transition vertices $\mathbb{N}_{ij}$ and neighbor vertices $N_{ij}$ separately leads to doubled data storage overhead on storing the data of $\mathbb{N}_{ij}\cap N_{ij}$. To avoid this issue, \sysname merges $\mathbb{N}_{ij}$ and $N_{ij}$ and maintains the data of $\mathbb{N}_{ij}\cup N_{ij}$ with a single data buffer. Additionally, \sysname regenerates the position indices and modifies the topology of each subgraph to ensure that the computation engine can read and write the merged data buffer directly.

{
\Paragraph{Computation engine}
\sysname’s computation engine is based on cuSparse and Pytorch \cite{pytorch_nips2019}, operating independently from the communication engine, as shown in Algorithm \ref{alg:exec_flow} (lines 8, 17, and 18). Following existing frameworks such as Sancus \cite{SANCUS_VLDB_2022} and DGL \cite{DGL_ARXIV_2019}, \sysname organizes the topology of each subgraph chunk into the compressed sparse row/column (CSR/CSC) formats. These subgraph chunks are abstracted as \texttt{block}s in the computation engine, facilitating GNN computations at each layer. Graph operations are implemented using cuSparse, while Pytorch serves as the backend for neural network computation. \sysname provides a GNN layer definition class with \texttt{\_\_init\_\_} and \texttt{forward} methods, enabling users to specify the model configuration and forward computation using built-in graph operations and Pytorch functions. Users also have the option to train their self-implemented GNN models in Pytorch or DGL by overloading these functions with their single-process codes. In case a different graph input format is used, users are required to convert the partitioned subgraphs into their preferred format in the preprocessing stage. The dataflow graph and autograd libraries of Pytorch are used for gradient computation, relieving users from explicitly managing the gradient calculations.
}

\section{Experimental Evaluation}
\label{sec:expr}

\subsection{Experimental Setup}
\label{sec:expr:setup}

\Paragraph{Environments} The multi-GPU experiments are conducted on a GPU server equipped with 4 AMD EPYC 7543 CPUs, 512GB DRAM, and 4 NVIDIA A100 (80GB) GPU. Each GPU is connected to a CPU via PCIe 4.0 link, and each CPU contains 128GB of local memory. The four GPUs are connected through NVLINK-3.0, providing 200GB/s inter-GPU bandwidth. The server runs Ubuntu 18.04 OS with GCC-7.5, CUDA 11.2 and PyTorch v1.9 backend \cite{pytorch_system}. 
The single-node CPU experiments are conducted on a Server contains two Intel Xeon 6246R CPU @3.40 GHz with a total of 32 cores and 768 GB of memory. The distributed CPU experiments are conducted on a 16-node Aliyun ECS cluster. Each node (ecs.r5.16xlarge instance) is equipped with 56 vCPUs and 512GB DRAM. The network bandwidth is 20 Gbps. All these machines run Ubuntu 20.04.

\begin{table}[!t]
\vspace{-0.1in}
	\caption{Dataset description. \revision{$|V|$, $|E|$ , $\#\mathbb{F}$, and $\#\mathbb{L}$ represent the number of vertices, edges, features, and labels, respectively.}}
	\vspace{-0.1in}
	\label{tab:Dataset}
	\centering
	\footnotesize
	{\renewcommand{\arraystretch}{1.2}
	\begin{tabular}{l r r c c c c c}
		\hline
		
		\hline
		{\textbf{Dataset}} &
		{\textbf{|V|}} &
		{\textbf{|E|}}  &
		{\textbf{\#$\mathbb{F}$}}&
		{\textbf{\#$\mathbb{L}$}}&
		{\textbf{Type}}\\
		\hline
		{reddit \cite{REDDIT_2017} (RDT)} & 0.23M & 114M &602&41& post-to-post\\		
		{ogbn-products \cite{OGB_DATASET} (OPT)}  &2.4M& 62M&100&47&co-purchasing \\
        {it-2004 \cite{IT_2004} (IT)}&	41M &1.2B&256&64& web graph\\
		{ogbn-paper \cite{OGB_DATASET} (OPR)}& 111M &1.6B&200&172&citation network\\
        {friendster} \cite{FS}(FDS)& 65.6M & 2.5B& 256& 64&social network\\
		\hline
		
	\end{tabular}
	}
	\vspace{-0.05in}
\end{table}

\Paragraph{Datasets and GNN algorithms}
Table \ref{tab:Dataset} presents the major parameters of the real-world graphs used in our experiments. For graphs without ground-truth properties (it-2004 and friendster), we use randomly generated features, labels, training (25\%), test (25\%) and validation (50\%) set division. We use two popular GNN models with different computation patterns, GCN \cite{GCN_ICLR_2017} has heavy-weight vertex computation and light-weight edge computation. GAT \cite{GAT_ICLR_2018} has heavy-weight edge computation and light-weight vertex computation. The hidden layer dimensions for reddit and ogbn-products are set to 256, while for it-2004, ogbn-paper, and friendster, they are set to 128. In our evaluation, the number of partitions is set to 4. Since reddit and ogbn-products are small, their partitions are not additionally split. Each partition of it-2004, ogbn-paper, and friendster is divided into 8, 32, and 32 (resp. 16, 64, 64) chunks in GCN (resp. GAT) training, respectively.


\Paragraph{The systems for comparison}
 We compare \sysname with three full-graph GNN systems: single-GPU DGL v0.9 \cite{DGL_ARXIV_2019}, multi-GPU-based Sancus \cite{SANCUS_VLDB_2022}, and CPU-based DistGNN \cite{DISTGNN_ARXIV_2021}, as well as a GPU-based mini-batch GNN system DistDGL\cite{DISTDGL_ARXIV_20}.
 \revision{In DistDGL, the fan-out of neighbor sampling per layer is set to 10, and the batch size is set to 1024.} We provide an in-memory version (\sysname-IM) that places all the training data in GPU to demonstrate the effectiveness of the GPU computation engine. \revision{We also provide a single-GPU \sysname with the inter-GPU communication disabled for comparison with DGL and single-CPU-based DistGNN.}

\begin{figure*}[ht]
	\centering
 	\vspace{-0.03in}
	\includegraphics[width=\textwidth]{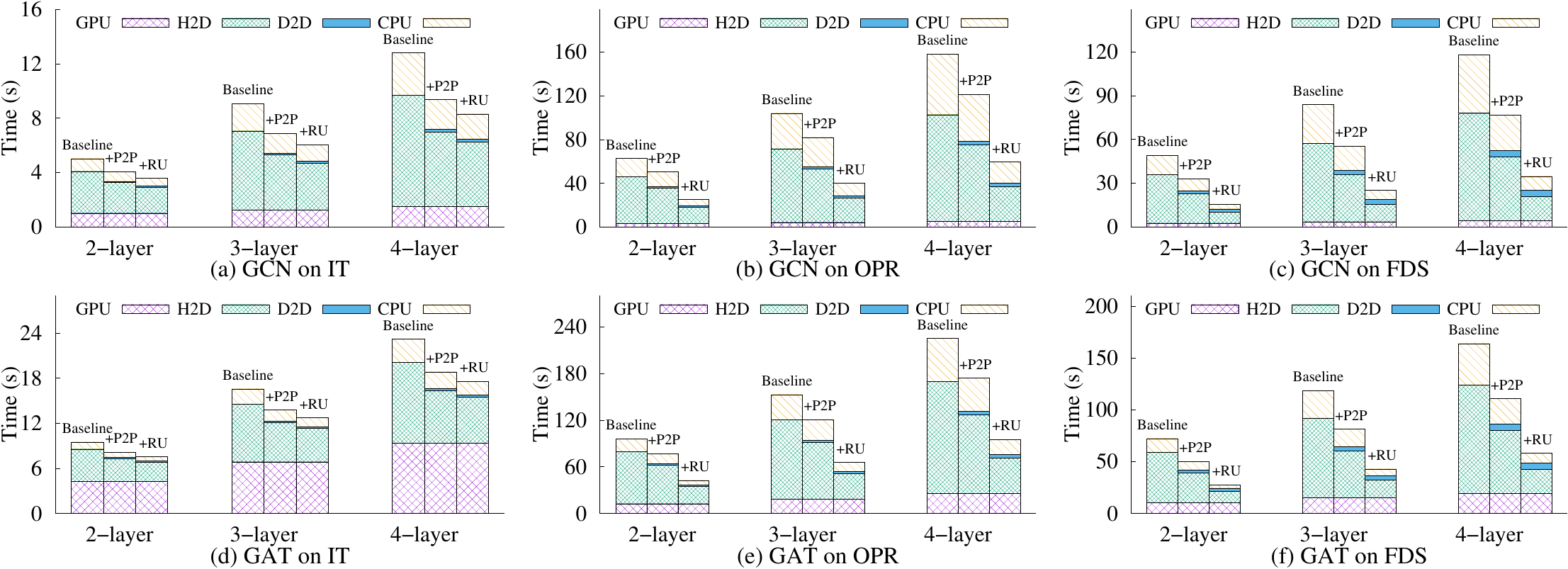}
 	\vspace{-0.12in}
	\caption{Performance breakdown of \sysnameb on GCN and GAT with different hidden layers, where `Baseline' for the baseline approach, `P2P' for the inter-GPU communication, and 'RU' for the intra-GPU data reusing. `GPU' represents the GPU computations, `H2D' represents the host-GPU communication, `D2D' represents the inter-GPU communication, and `CPU' represents the CPU-based gradient accumulation.}
	\label{fig:sec5:perf}
 \vspace{-0.05in}
\end{figure*}

{
\Paragraph{Comparison with CPU data offloading techniques in DNN training} Certain DNN frameworks also employ CPU data offloading to mitigate GPU memory overhead. DeepSpeed \cite{ZERO_SC_2020, ZEROOFFLOAD_ATC_2021, ZEROINFINITY_SC_2021} is a representative system that stores model parameters in CPU and offload computation to GPUs.
However, these frameworks are designed for DNN training with large models and lack GNN-specific consideration, limiting their effectiveness in supporting GNN training (Section \ref{sec:relate}). To illustrate this, we compare \sysname with DeepSpeed \cite{DeepSpeed} in Section \ref{sec:expr}.3. Since DeepSpeed does not support GNN training, we implement its data offloading method in \sysname as our baseline (denoted by \textbf{Baseline} in Figure \ref{fig:sec5:perf}), which transfers the neighbor data for each subgraph entirely. The host-GPU on-demand access optimization (Section \ref{sec3:impl_comm}) is enabled in the baseline to enhance CPU-GPU communication. Both DeepSpeed and \sysname employ recomputation-based training \cite{SUBMEM_ARXIV_2016}. For a fair comparison, we enable recomputation-cache hybrid intermediate data management (Section \ref{sec3:rc}) in both frameworks, even though DeepSpeed does not have this optimization.

}


\begin{figure}[t]
	\centering
	\includegraphics[width=0.75\linewidth]{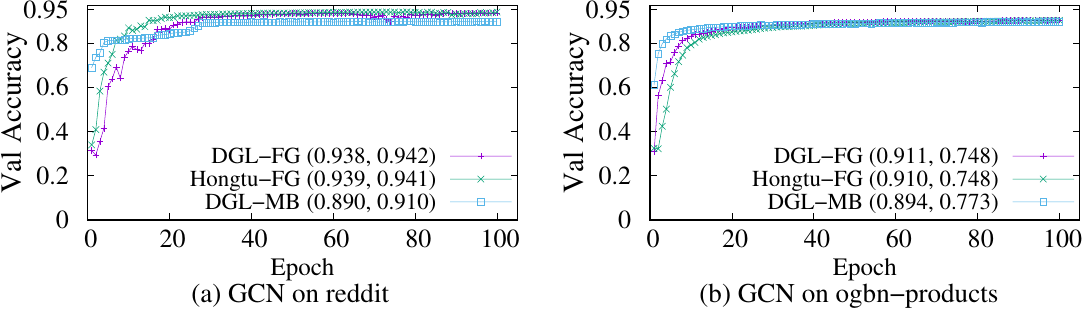}
    \vspace{-0.05in}
	\caption{Validation accuracy curves of DGL (full-graph), DistDGL (mini-batch) and \sysnameb for GCN with 100 epochs. The values in (\#val, \#test) are the final validation and test accuracy, respectively.}
	\label{fig:sec5:acc}
    \vspace{-0.05in}
\end{figure}

{
 \Paragraph{Accuracy and evaluation metric}
Full-graph GNN can achieve theoretical accuracy in \sysname because its training semantic is not changed. Figure \ref{fig:sec5:acc} shows the validation and test accuracy of \sysname and DGL in full-graph training for a GCN model. After 100 epochs, the validation accuracy reached a stable state, \sysname and DGL-FG almost achieved the same validation and test accuracy. Since the accuracy gap between DGL and \sysname is almost negligible, we report the per-epoch runtime, i.e., the time to conduct a forward and backward pass over the entire graph. Shorter per-epoch time indicates better time-to-accuracy performance, and all the results are averaged over 20 epochs to ensure consistency. In comparison with mini-batch training-based DGL, \sysname achieves higher test and validation accuracy on reddit, while mini-batch training performs better on ogbn-products. Both the mini-batch and full-graph training methods possess distinct merits. However, assessing their effectiveness requires a comprehensive analysis and consideration of various factors, including batch size, sampling fan-out, and characteristics of input graphs, which is out of the scope of this work.
}


\begin{table}[h]\vspace{-0.1in}
	\caption{Comparison with DGL and DistGNN on two small datasets.}
	\vspace{-0.05in}
	\label{tab:sec5:single}
	\centering
	\footnotesize
	{\renewcommand{\arraystretch}{1.0}
	\begin{tabular}{c |r l l| l l}
		\hline
		
		\hline
		\multirow{2}*{ Layers}&\multirow{2}*{\textbf{System}} &
		\multicolumn{2}{c}{ Runtime of GCN (s)} & \multicolumn{2}{c}{ Runtime of GAT (s)}\\
		\cline{3-4} \cline{5-6}
		&&{\textbf{RDT}} &
		{\textbf{OPT}}  &
		{\textbf{RDT}}&
		{\textbf{OPT}}\\
		\hline

        \hline
		\multirow{4}*{2}&{DistGNN} & 4.2&10.1 & 40.7&49.9\\
		&{DGL}& 0.19 (21$\times$) & 0.27 (37$\times$) &0.86 (47$\times$)& 1.22 (41$\times$)\\
        &\sysname-IM&0.20 (21$\times$)&0.32 (31$\times$)&0.77 (53$\times$)&1.14 (43$\times$)\\
		&{\sysname} & 0.33 (12$\times$) &0.84 (12$\times$) & 1.15 (35$\times$) & 1.93 (26$\times$)\\
		\hline
    
        \hline
		\multirow{4}*{4}&{DistGNN} & 7.78 &22.9 & 77.9&220.8\\
		&{DGL}& 0.39 (20$\times$) & 0.82 (28$\times$) &1.35 (58$\times$)& 2.19 (101$\times$)\\
        &\sysname-IM&0.39 (20$\times$)&0.81 (28$\times$)&1.21 (64$\times$)&2.01 (109 $\times$)\\
  
		&{\sysname} & 0.62 (13$\times$) &2.09 (11$\times$) & 2.21 (35$\times$) & 3.82 (58$\times$)\\
		\hline

        \hline
		\multirow{4}*{8}&{DistGNN} & 15.1 &46.2 & 148.4&418.6\\
		&{DGL}& 0.78 (19$\times$) & 1.92 (24$\times$) &2.77 (54$\times$)& OOM\\
        &\sysname-IM&0.69 (22$\times$)&1.76 (13$\times$)&2.43 (26$\times$) &OOM\\
		&{\sysname} & 1.15 (13$\times$) &4.14 (11$\times$) & 4.50 (32$\times$) & 8.23 (50$\times$)\\
		\hline
  
		\hline
	\end{tabular}
 }
\end{table}

\subsection{Overall Comparison}

First, we compare \sysname with single-CPU and single-GPU systems on small graphs to show the efficiency of GPU computation engine. Then we compare \sysname with multi-GPU systems on all graphs to evaluate its processing scale with limited GPU resources. Finally, we compare \sysname with a distributed CPU system on large graphs, evaluating its efficiency and low monetary cost.

\Paragraph{Comparison with single-GPU and single-CPU systems}
We compare \sysname and \sysname-IM with DGL \cite{DISTDGL_ARXIV_20} and single-CPU DistGNN by running GCN and GAT on the two small graphs (reddit and ogbn-products). Table \ref{tab:sec5:single} shows the runtime results and the speedups normalized to DistGNN. We observe that all three GPU-based solutions achieve more than one order-of-magnitude speedup over the CPU-based solution. \sysname-IM achieves performance similar to, or slightly better than, DGL. \sysname is 1.3$\times$-3.8$\times$ slower than DGL due to additional overhead on host-GPU communication and CPU-based gradient accumulation. Although the performance is slightly behind, only \sysname is capable of training complex GNN models with large-scale intermediate data (e.g., GAT).


\begin{table}[t]
	\vspace{-0.07in}
	\caption{Comparison with Multi-GPU systems on 4 A100 GPUs.}
	\vspace{-0.1in}
	\label{tab:sec5:multi}
	\centering
	\footnotesize
	{\renewcommand{\arraystretch}{1.0}
	\begin{tabular}{c|l |l l l l l}
		\hline
		
		\hline
		\multirow{2}*{ Layers}&\multirow{2}*{System}
		&\multicolumn{5}{c}{ Runtime of GCN (s)} \\
		\cline{3-7}
		&&{\textbf{RDT}} &
		{\textbf{OPT}}  &
		{\textbf{IT }}&
        {\textbf{OPR}}&
		{\textbf{FDS}}\\
		\hline

        \hline
        \multirow{4}*{2/2}
		&Sancus & 0.09&0.37 & OOM&OOM&OOM\\
         &\sysname-IM&0.09(1.0$\times$)&0.29(1.27$\times$)&OOM&OOM& OOM\\
        &\sysname  &0.13(0.72$\times$) &0.45(0.81$\times$) &3.6 &25.1&15.5\\
        \cline{2-7}
        &DistDGL & 0.61(0.15$\times$)&0.58(0.64$\times$) & 33.8&1.95 & 52.3\\
		\hline
    
        \hline
        \multirow{4}*{4/3}
		&Sancus & 0.15 & 0.65 &OOM&OOM&OOM\\
  
        &\sysname-IM  &0.16(0.93$\times$)&0.55(1.18$\times$)&OOM&OOM&OOM\\
        &\sysname  &0.23(0.63$\times$) &0.95(0.68$\times$) &6.0&40.0&25.1\\
        
        \cline{2-7}
        &DistDGL & 1.49(0.10$\times$)&4.13(0.16$\times$) & 281.5&6.95&397.2\\
        \hline

        \hline
        \multirow{4}*{8/4}
		&Sancus  & 0.27 & 1.16& OOM&OOM&OOM\\
        &\sysname-IM  &0.25(1.08$\times$)&1.04(1.15$\times$)&OOM&OOM&OOM\\
                &\sysname  & 0.45(0.65$\times$) &2.03(0.57$\times$) & 8.3& 59.4 &34.6\\
                \cline{2-7}
        &DistDGL & 22.4(0.01$\times$) & OOM & OOM&19.2&OOM\\

        \hline
    
		\hline
	\end{tabular}
 }\vspace{-0.1in}
\end{table}


\Paragraph{Comparison with multi-GPU system} We compare \sysname with Sancus \cite{SANCUS_VLDB_2022} and \revision{DistDGL \cite{DISTDGL_ARXIV_20}} by running GCN on all five graphs. For the two small graphs, we employ the model configurations with 2, 4, and 8 layers, while for the three large graphs, we employ model configurations with 2, 3, and 4 layers. The results are reported in Table \ref{tab:sec5:multi}.  In comparison with Sancus, \sysname-IM delivers comparable performance to Sancus and is 1.2$\times$-1.9$\times$ faster than \sysname on the two small graphs. However, both Sancus and \sysname-IM run out of memory on the three large graphs. In contrast, \sysname can effectively process them. 
\revision{
DistDGL runs out of memory on ogbn-products, it-2004, and friendster when configured with 8, 4, and 4 layers, respectively. Furthermore, in cases where DistDGL successfully runs, the runtime exhibits exponential growth as the number of layers increases. These challenges arise from the neighbor explosion problem \cite{ROC}, where the computation and memory requirements for mini-batch GNN training increase exponentially with the number of layers. On successfully runs, \sysname outperforms DistDGL on reddit, ogbn-products, it-2004, and friendster. On ogbn-paper, DistDGL achieves a better performance due to its usage of only 1.2M vertices (1.1\%) for the training, resulting in significantly lower computation volume compared to \sysname. In summary, \sysname exhibits advantages when training deep GNNs or when the input graph includes a large proportion of training vertices. However, when the training set and the number of model layers are small, DistDGL still holds certain advantages.
}

\begin{table}[t]
\vspace{-0.0in}
	\caption{Comparison with DistGNN on a 16-node ECS cluster.}
	\vspace{-0.05in}
	\label{tab:cpu}
	\centering
	\footnotesize
	{\renewcommand{\arraystretch}{1.0}
	\begin{tabular}{c |c l l| l l}
		\hline
		
		\hline
		\multirow{2}*{Layers}&\multirow{2}*{{Dataset}} &
		\multicolumn{2}{c}{Runtime of GCN (s)} & \multicolumn{2}{c}{Runtime of GAT (s)}\\
		\cline{3-4} \cline{5-6}
		&&{\textbf{DistGNN}} &
		{\textbf{\sysnameb}}  &
		{\textbf{DistGNN}}&
		{\textbf{\sysnameb}}\\
		\hline

        \hline
		\multirow{3}*{2}&{IT} & 38.9 &3.6 (10.8$\times$) & 151.3&7.5 (20.2$\times$)\\
        &{OPR} & 213.5& 25.1 (8.5$\times$)& OOM&42.4\\
		
        &{FDS}  & 183.0& 15.5 (11.8$\times$)& OOM&27.5\\
		\hline
    
        \hline
		\multirow{3}*{3}&{IT} & 59.5&6.0 (9.9$\times$) &OOM &12.7\\
        &{OPR} & 312.6 & 40.0 (7.8$\times$)&OOM &65.6\\
		&{FDS}  &277.5 & 25.1 (11.1$\times$)&OOM &42.8\\
		\hline

        \hline
		\multirow{3}*{4}&{IT} &85.7 & 8.3 (10.3$\times$)&OOM &17.5\\
         &{OPR}  &OOM & 59.4& OOM&95.0\\
		&{FDS}  & 369.4& 34.6 (10.7$\times$)& OOM&58.0\\
		\hline
  
		\hline
	\end{tabular}
 }
\end{table}

\Paragraph{Comparison with distributed-CPU system}
We compare \sysname with DistGNN \cite{SANCUS_VLDB_2022} by running GCN and GAT on the three large graphs. The results are reported in Table \ref{tab:sec5:multi}. DistGNN runs out-of-memory for 4-layer GCN on ogbn-paper and all GAT workloads except the two-layer GAT training on it-2004. We can observe that training large-scale GNNs remains challenging, even with extensive host memory provided by multiple CPU nodes. Besides large-scale vertex and intermediate data, DistGNN also needs to maintain the data of neighbor replicas and communication buffers for distributed processing.
In other cases, \sysname outperforms CPU-based solution. On average, \sysname achieves 10.1$\times$ and 20.2$\times$ speedups over DistGNN for GCN and GAT models, respectively. Moreover, the per-hour monetary cost for 16 CPU nodes (ecs.r5.16xlarge, 5.24 \texttt{USD} per-nodes) is 4.16 times higher than that for a 4$\times$A100 GPU node (ecs.gn7e-c16g1.16xlarge, 20.14 \texttt{USD} per-node), which has a similar configuration to our private GPU server. Therefore, \sysname provides a cost-effective solution for processing large-scale GNNs.

\subsection{Communication Reduction Analysis}
\label{sec:expr:gain}

\begin{table}

	\caption{The proportion of the two types of duplication access on the three billion-scale graphs.}
	\vspace{-0.11in}
	\label{tab:sec3:duplication}
	\centering
	\footnotesize
	{\renewcommand{\arraystretch}{1.2}
	\begin{tabular}{l c c c c c c c}
		\hline
		
		\hline
		\multirow{1}*{\textbf{Dataset}} &
		\multirow{1}*{\textbf{Chunks}}&
        \multirow{1}*{${\textbf{V}_{ori}}$}   &
		\multirow{1}*{$(\textbf{V}_{ori}-\textbf{V}_{+p2p})$}&
		\multirow{1}*{$(\textbf{V}_{+p2p}-\textbf{V}_{+ru})$}\\
		\hline
        {it-2004} &32&1.6&0.26 (16.2\%)&0.15 (9.2\%)\\
		{ogbn-paper} &128&8.5&0.77 (9.0\%)& 4.1 (48.3\%)\\
        {friendster}&128&10.7&2.50 (23.3\%) &5.09 (47.6\%)\\
		\hline

	\end{tabular}
	}
\end{table}

We enable inter-GPU and intra-GPU communication deduplication one-by-one to reveal how much \sysname can benefit from each of them. Table \ref{tab:sec3:duplication} illustrates the communication reduction volume normalized to the number of vertices ($|V|$).The results show that these two optimizations reduce host-GPU communication by 25\%-71\% on the three graphs. Although it-2004 originally has less redundant communication ($0.6$ times $|V|$), our proposed method still reduces 68\% of the total redundant transfers (from $0.6|V|$ to $0.2|V|$). Ogbn-paper benefits more from intra-GPU deduplication due to its co-author graph structure and exhibits good locality. 

To demonstrate the practical improvement of communication deduplication, we conducted experiments on GCN and GAT models with 2-, 3-, and 4-layer configurations on the three large graphs.  We start from the baseline approach (\textbf{Baseline}) that transfers the neighbor data for each subgraph entirely, then enable inter-GPU (denoted by \textbf{+P2P}) and intra-GPU communication deduplication (denoted by \textbf{+RU}) one-by-one. Figure \ref{fig:sec5:perf} reports the results.  Even with on-demand access optimization (Section \ref{sec3:impl_comm}), the performance of the baseline approach remains inferior, because it suffers from the duplicated neighbor data communication and cross-partition remote host-GPU data access. The inter-GPU data sharing reduces communication time (including host-GPU communication and inter-GPU communication ) by 23\%- 26\%, 23\%-27\%, and 39\%- 42\%, on the three graphs respectively. The reduction in transfer time is greater than the reduction in transfer volume because it eliminates the remote neighbor access across CPUs. The intra-GPU data reusing further reduces transfer time by  9\%-12\%, 39\%-42\%, and 36\%-37\% for the three graphs, respectively. Overall, \sysname that uses deduplicated communication can achieve speedups ranging from 1.3$\times$ to 3.4$\times$ compared to the baseline approach.

{
 
\begin{table}[t]
    \vspace{-0.05in}
	\caption{Analysis of cost of communication deduplication.}
	\vspace{-0.11in}
	\label{tab:cost-gain}
	\footnotesize
	{\renewcommand{\arraystretch}{1.2}
	\begin{tabular}{l r r r}
		\hline
		
		\hline
		\multirow{2}*{\textbf{Engine}} &
		\multicolumn{3}{c}{Runtime of 100 epochs on a 2-layer GCN(s)}\\
		\cline{2-4}
		&{\textbf{it-2004}} &
		{\textbf{ogbn-paper}}  &
		{\textbf{friendster}}\\
		\hline
		{\textbf{\sysname} w/o CD}& 502.8 & 6260.2 & 4907.5\\
		\hline
		{\textbf{\sysname} w/ CD}&359.6 &2513.0& 1554.1\\
	    {{Preprocessing}}& +4.5 &+33.9 & +22.7\\
		\hline

		\hline
	\end{tabular}
	}
\end{table}

\Paragraph{Overhead of communication deduplicaton}
As communication deduplication has the cost to preprocess the input graph after graph partitioning, we evaluate the overhead and show in Table \ref{tab:cost-gain}, where the preprocessing time is denoted as "Preprocessing". We compare it to the execution time of running GCN for 100 epochs with and without communication deduplication (CD). We observe that communication deduplication brings up to 1.5\% overhead into \sysname while significantly improving performance over the baseline. The low overhead of communication deduplication comes from two folds. First, the preprocessing uses a heuristic design and is executed in parallel. Second, as full-graph GNN training follows the same execution pattern in different layers, the preprocessing only needs to be performed once.
}

\vspace{-0.04in}
\subsection{Performance Breakdown}
\vspace{-0.01in}
We provide a performance breakdown to analyze the time consumption of different components, including the host-GPU communication (H2D), inter-GPU communication (D2D), GPU-based computation (GPU), and CPU-based gradient accumulation (CPU). Figure \ref{fig:sec5:perf} shows the experimental results. The GPU computation time varies among different GNNs due to their varying computation complexities. In GCN with simple arithmetic edge computation, GPU computation accounts for 10\%-14\% of the overall runtime. In contrast, in GAT with parameterized edge computation, the GPU computation time is 4.5 times longer than that of GCN and accounts for 54\%, 28\%, and 35\% of the total runtime. The communication (H2D+D2D) time varies among different GNNs. GCN benefits from recomputation-caching-hybrid training, reducing its communication time by 21\%-29\% compared to GAT training. Overall, the communication time accounts for 58\%-61\% and 36\%-50\% of the overall runtime on GCN and GAT, respectively. As \sysname utilizes CPUs to accumulate the neighbor gradients, the CPU computation time is proportional to the volume of transferred neighbors, which accounts for 8\% to 30\% of the overall runtime.

\subsection{Sensitivity Study}

\Paragraph{Performance with varying layers} Since the computation pattern is exactly the same, communication deduplication is equally effective for all GNN layers. Therefore, increasing the number of layers will not decrease the optimization effect. As shown in Figure \ref{fig:sec5:perf}. \sysname achieves 1.4$\times$-1.5$\times$, 2.5$\times$-2.7$\times$, 3.2$\times$-3.4$\times$, 1.3$\times$-1.3$\times$, 2.3$\times$-2.4$\times$, and 2.6$\times$-2.8$\times$ speedups over the vanilla approach under different layer configurations. The optimization effect is stable.

\begin{figure}[t]
	\centering
	\includegraphics[width=0.7\linewidth]{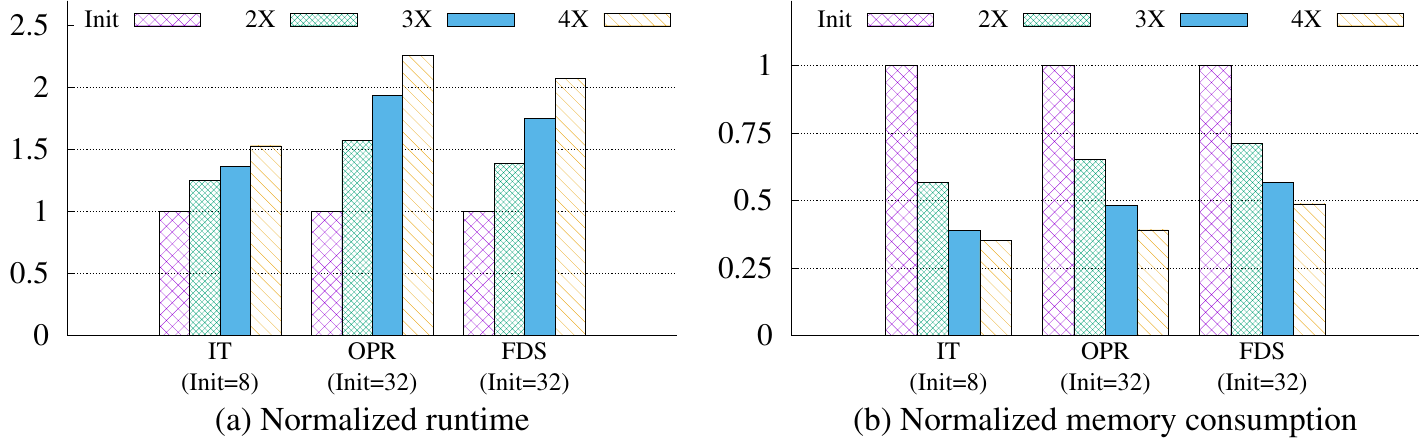}
 \vspace{-0.1in}
	\caption{Runtime and memory analysis of \sysnameb with different chunks. We run GCN on each graph from an initial chunk size and increase the chunk size to 2$\times$, 3$\times$, and 4$\times$.}
	\label{fig:sec5:varying}
 \vspace{-0.1in}
\end{figure}

\begin{figure}[t]
	\centering
	\includegraphics[width=0.65\linewidth]{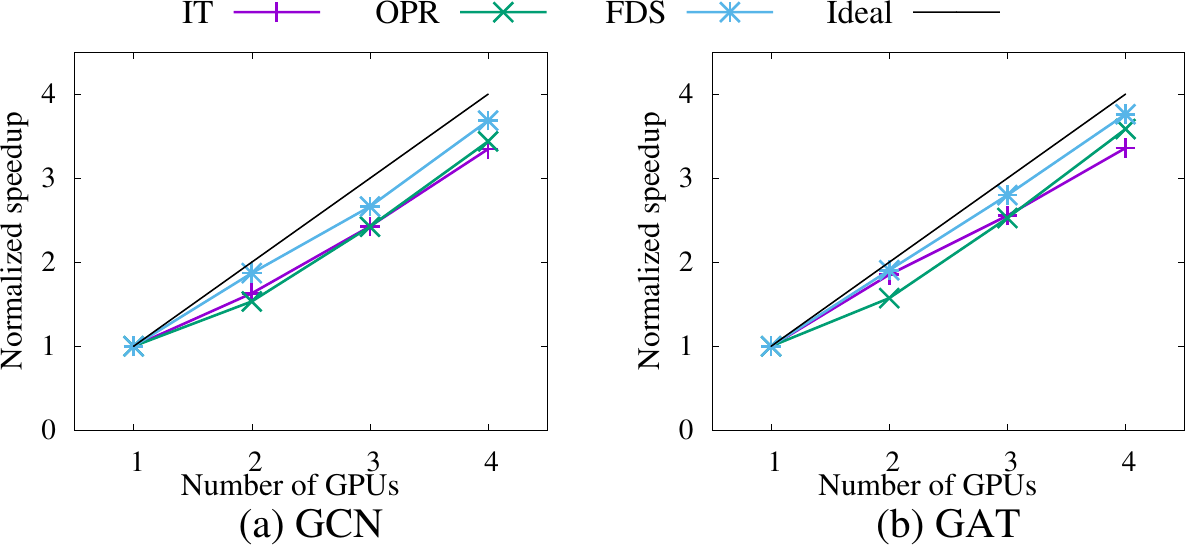}
    \vspace{-0.14in}
	\caption{Scaling perf. when varying GPU number from 1 to 4.}
	\label{fig:sec5:scale}
    \vspace{-0.15in}
\end{figure}

\Paragraph{Performance with varying chunks} The chunk size in \sysname is a configurable parameter that controls the memory consumption of training data. However, increasing the chunk size also leads to increased duplicated neighbors, which subsequently increases the volume of host-GPU communication. To evaluate the impact of chunk size, we run GCN on three large graphs, starting from the initial chunk size and increasing it by a factor of 2, 3, and 4. The experimental results in Figure \ref{fig:sec5:varying} show that as the chunk size increases by 4$\times$, the memory consumption decreases by 51\%-65\%, and the runtime increases by 1.5$\times$ to 2.2$\times$. The increase in runtime is either linear or sublinear, and is proportional to the decrease in memory consumption. Additionally, the performance of \sysname can be improved by using GPUs with larger memory capacity, although our approach can be adapted to GPUs of different grades.

\subsection{Scalability of \sysname}
We evaluate the scalability of \sysname by varying the number of GPUs used in training.  Figure \ref{fig:sec5:scale} shows the normalized speedups of GCN and GAT training on it-2004, ogbn-paper, and friendster. The execution time of \sysname is reduced when using more GPUs. Specifically, when the number of GPUs increases from 1 to 4, \sysname achieves 3.3-3.7$\times$ speedups for GCN training and 3.4-3.8$\times$ speedups for GAT training. The speedups from 1 to 2 sockets is lower than that from 2 to 4 sockets because we do not have enough CPU memory to enable the NUMA-aware vertex data allocation. When using two or fewer GPUs, we must use the memory from all sockets, resulting in remote memory access overhead.

\section{Related Work}

\label{sec:relate}
{
As the size of DNN models increases, the traditional data parallel (DP) training method, which replicates model parameters across all training processes, faces scalability issues due to the increasing memory consumption \cite{TENSOR_OSDI_2016,PIPEDREAM_SOSP_2019,MODELPARALLEL_EUROSYS_2019}. To address this, various parallel training methods have been proposed, including model parallelism \cite{MODELPARALLEL_NIPS_2012,MODELPARALLEL_EUROSYS_2019}, pipeline parallelism \cite{GPIPE_NIPS_2019,PIPEDREAM_SOSP_2019,MEPIPE_ICML_2021}, partitioned data parallelism \cite{ZERO_SC_2020} (which partitions model states among data-parallel workers to eliminate memory redundancy), and 3D parallelism \cite{DeepSpeed} (which combines model, pipeline, and data parallelism to leverage the aggregated GPU memory of a cluster). However, the scalability of these frameworks remains constrained by the available GPU resources.
To address the limitation of GPU memory, DeepSpeed \cite{DeepSpeed} incorporates CPU-based data offloading techniques \cite{ZERO_SC_2020,ZEROOFFLOAD_ATC_2021,ZEROINFINITY_SC_2021}. It achieves this by partitioning the model state into smaller slices and storing them in CPU DRAM or NVM. During training, the required slices are loaded into the GPUs sequentially as they are accessed. While DeepSpeed and our design share similarities in offloading memory-intensive data to CPU memory, they differ in their optimization objectives. DeepSpeed primarily focuses on DNNs with large model parameters. In DNNs, model parameters consist of dense matrices that can be partitioned into disjoint slices without interdependencies. These slices can be efficiently communicated between the CPU and GPUs due to their regular data access patterns. In contrast, \sysname is tailored for GNN training, where the memory overhead primarily arises from vertex data, and the model data typically have small sizes. Due to the inherent complexity of graph structures, vertex data are randomly distributed and duplicated across partitions, resulting in irregular and increased host-GPU data communication. DeepSpeed's approach does not adequately address these challenges. However, \sysname effectively resolves this problem through its deduplicated communication framework.

}
\section{Conclusion}
We present \sysname, a scalable and efficient system for training full-graph GNNs on limited GPU memory. Our system leverages two key components to achieve its performance, including a memory-efficient GNN training framework that combines the partition-based GNN training and recomputation-cache-hybrid intermediate data management, a deduplicated communication framework that converts the redundant host communication for duplicated neighbors to inter-GPU and intra-GPU data access. Our experiments demonstrate that \sysname can efficiently train on billion-scale graphs using just 4 GPUs 
by fully utilizing CPU, GPU, and interconnects.



\begin{acks}
We thank the anonymous reviewers for their constructive comments and suggestions. This research/project is supported by the National Research Foundation, Singapore under its AI Singapore Programme (AISG Award No: AISG2-TC-2021-002), the Ministry of Education AcRF Tier 2 grant (No. MOE-000242-00/MOE-000242-01), a grant from NUS Advanced Research and Technology Innovation Centre (ARTIC), and Google South \& Southeast Asia Research Award 2022.
\end{acks}


\bibliographystyle{ACM-Reference-Format}

\end{document}